\documentclass{aa}
\usepackage{graphicx,epsfig,rotating}
\def\Agata{R\' o\. za\' nska~}

\begin{document}

               
\title{The structure and radiation spectra of illuminated accretion 
 disks in AGN.\ II. Flare/spot model of X-ray variability}

\author{B.~Czerny,\inst{\!1}
        A.~R\'o\.za\'nska,\inst{\!1}
        M.~Dov\v{c}iak,\inst{\!2,3}
        V.~Karas\inst{2,3}
        \and
        A.-M.~Dumont\inst{4}}
\offprints{A. R\'o\.za\'nska (agata@camk.edu.pl)}
\institute{
$^1$~Copernicus Astronomical Center, Bartycka 18, 00-716 Warsaw, Poland\\
$^2$~Astronomical Institute, Academy of Sciences of the Czech Republic,
     Prague, Czech Republic\\
$^3$~Faculty of Mathematics and Physics, Charles University, 
     Prague, Czech Republic\\
$^4$~Observatoire de Paris-Meudon, DAEC, Meudon, France}

\authorrunning{B.~Czerny et al.}
\titlerunning{Flare/spot model of AGN X-ray variability}
\date{Received ...; accepted ...}

\abstract{We discuss a model of X-ray variability of active galactic 
nuclei (AGN). 
We consider multiple spots which originate on the surface of an 
accretion disk following
intense irradiation by coronal flares. 
The spots move with the disk around the central black
hole and eventually decay while new spots continuously emerge. 
We construct time sequences 
of the spectra of the spotted disk and compute the corresponding 
energy-dependent 
fractional variability amplitude. 
We explore the dependence on the disk inclination and other model parameters. 
AGN seen at higher inclination with respect to the observer, such as 
Seyfert~2 galaxies, are expected to have 
fractional variability amplitude of the direct emission 
by a factor of a few higher 
than objects seen face on, such as the Seyfert~1s. 
}
\maketitle

\section{Introduction}
Broad band spectra of active galactic nuclei show the presence of several
components. The two principle contributions are: 
(i)~the Big Blue Bump extending from optical/UV band to (sometimes) 
soft X-ray band, and (ii)~the hard X-ray power law. The Big Blue 
Bump emission is conventionally interpreted as originating from an
accretion disk (Czerny \& Elvis 1987; Koratkar \& Blaes 1999; however,
see e.g.\ Collin et al.\ 2002 for discussion of problems encountered by this 
scheme). On the other hand, the nature and geometry of the region 
responsible for the hard X-ray emission is still under discussion. Several 
models were proposed, the most popular ideas being a hot extended corona
overlaying a relatively cold disk (e.g. Czerny \& Elvis 1987, Haardt \& Maraschi 
1991, \Agata \& Czerny 2000, Liu et al. 2002, Merloni 2003), 
an inner hot flow (e.g. Ichimaru 1977, Narayan \& Yi 1994, Narayan et al. 2002) 
the lamp-post model (where a point-like 
X-ray source is located at a specified height above the disk, e.g. Henri \&
Pelletier 1991, Malzac et al. 1998), and the model of 
multiple hot flares produced via magnetic field reconnections 
(Galeev et al. 1979 and later papers). The field was reviewed, for example, by 
Leighly (1999), Collin (2001), and  Poutanen (1998) and Done et al. (2002) 
(in the context of X-ray 
binaries). Notice that these scenarios are not completely disparate
and there may be a certain overlap between them. 
In all models the primary source of X-rays should be strongly 
variable.

Indeed, AGN are variable in the X-ray band (e.g.\ Lawrence et al.\ 1987; 
Taylor et al.\ 1993) but the origin of variability cannot be examined
directly, because the relevant central regions still remain unresolved 
to direct imaging. However, studies of the X-ray spectral variability
offer a direct insight into the structure of accretion flows onto a
central black hole, which is assumed to power AGN and determine their
spectra.

In the flare model, sudden dissipation occurs in very localized regions 
above the disk surface (coronal loops, in analogy with the solar corona). 
In this case the local irradiation flux can be orders of 
magnitude higher then the steady level of the disk flux itself, but the 
irradiation does not last very long and only a small fraction of the disk 
surface is irradiated at every moment.

Such arrangement of X-ray emitting region has been suggested for the first time
by Galeev et al.\ (1979), and it was subsequently developed in many papers
(e.g.\ Abramowicz et al.\ 1991; Haardt et al.\ 1994; van Oss et al.\ 1993; 
Poutanen \& Fabian 1999; \.Zycki 2002). 
The development of a flare leads to a burst of `primary emission'
as well as to the formation of a hot spot underlying the flare where 
roughly half of the X-ray flux is reprocessed by the disk. Irradiation of the 
disk surface in hydrostatic equilibrium leads to the formation of strongly 
stratified medium -- a hot fully ionized skin 
covering a cooler, more neutral zone -- and so the X-ray spectrum resulting 
from 
reprocessing should contain signatures that are characteristic for 
multi-temperature
gas (Nayakshin et al.\ 2000; Ballantyne et al.\ 2001; 
R\'o\.za\'nska et al.\ 2002).  
The irradiating flux is locally very large, exceeding considerably 
the stationary 
energy flux which is dissipated inside the disk (e.g.\ Nayakshin 2000; 
Ballantyne et al.\ 2001). The resulting ionized skin is relatively
optically thick (Collin et al.\ 2003).

The flare/spot model is a possible (although not unique) explanation of 
narrow emission features, which have been reported in $\sim5$--$6$~keV 
X-ray spectra of several AGNs and interpreted in terms of in terms of 
localized iron-line emission (Turner et al. 2002, 2004; Guainazzi 2003; 
Yaqoob et al. 2003; Dovciak et al. 2004).

In the present paper we test the flare/spot model by analyzing its 
predictions for the fractional variability amplitude in the X-ray band.
We model the local spot/flare spectrum as a sum of flare (primary) emission 
of a power law shape, and a spot (reflected) emission. The spot emission is
determined by using the coupled {\sc titan/noar} codes (Dumont et al.\ 2000) 
to solve the radiative transfer and the code of R\'o\.za\'nska et al.\ (1999) 
to calculate the hydrostatic equilibrium. We assume random distribution of 
spots and flares across the disk and we account for their motion during the 
time-integrated observation. We consider both non-rotating
(Schwarzschild) and rapidly rotating (Kerr) black holes, and we apply 
general relativity corrections using the {\sc ky}  code of 
Dov\v{c}iak et al.\ (2003).

In our approach, a sequence of solutions is specified by fixing
statistical properties which the flare/spot distribution is required 
to obey. Given a particular solution, the corresponding energy-dependent
fractional variability amplitude $F_{\rm{}var}$ is calculated.
The model is described in Section~\ref{sect:model}. Results are 
given in Section~\ref{sect:results}. The relevance of the 
model for explaining spectrum and variability of a typical 
Seyfert~1 galaxy is discussed in Section~\ref{sect:discussion}.

\section{The model}
\label{sect:model}

\subsection{Intrinsic spectrum of the spot}
\label{sect:local_spot}
We model the X-ray spectrum from an irradiated accretion disk around a 
massive black hole, assuming a power law profile of the flare 
primary emission and computing the radiation spectrum from a hot spot 
underlying the compact flare. The spot emission goes directly
to the observer without any further secondary Comptonization 
within the flare (we neglect size of the flare itself) 
and we adopt plane parallel symmetry for local computations.

Reprocessing of the incident flux is computed with the code 
{\sc titan} (Dumont et al.\ 2000). The radiative transfer is 
solved both for the continuum and for lines, because the escape 
probability approach does not give reliable results for the lines
(Dumont et al.\ 2003). The Comptonization is taken into account by 
coupling with the Monte Carlo code {\sc noar} (also Dumont et al.\ 2000). 

The vertical structure of the underlying disk is calculated with the 
code of R\'o\.za\'nska et al. (1999). As our principal model, we consider 
an irradiated disk in hydrostatic equilibrium, which requires iterations 
between the disk structure and the radiation transfer computations. 
These iterations are carried out as described by R\'o\.za\'nska
et al.\ (2002). The computations of equilibrium are extremely time-consuming
because of coupling between the equations of radiation 
transfer for the lines and for continuum. Therefore, in the present paper
we performed computations at a certain representative radius in the disk 
($R=18 R_{\rm g}$ in terms of gravitational radii, $R_{\rm g}{\equiv}GM/c^2$) 
with the aim of obtaining a typical reflection spectrum, $F_{\rm loc}(E)$. 
Local reflection spectra at other radii were obtained with the help of 
scaling. 

\subsection{Assumptions about the flare -- spot connection}
\label{sect:flares}
We intend to represent a typical observation of an AGN, and therefore we
choose as model parameters preferentially those properties which are
connected as close as possible to observations. We define a very simple
model (in fact, too simple to explain the origin of
power spectra themselves). The model takes into account the basic
observational properties, such as mean luminosity and
luminosity dispersion. This approach is complementary to  the papers
devoted to the development of an isolated flare (e.g. B\"{o}ttcher et al.\
2003).

Fundamental parameters of the model are the mass of the black hole, $M$, 
and the mean X-ray luminosity, $L_{\rm{}X}$. This X-ray luminosity 
is assumed to originate from $n_{\rm{}mean}$ flares per second,
on average. 
The actual number of flares fluctuates at any given moment 
around this mean value according to Poisson's distribution.
Flares are assumed to corotate with the underlying disk.

A fraction of radiation from each flare reaches the observer 
as the direct component. On the other hand, a part of flare photons 
is reprocessed by the disk surface, producing the reflection
component.

In the model, a flare occurs at height $h$ above the disc, it
irradiates the disk surface and creates a hot spot.
Irradiation is the largest and strictly perpendicular just
below the flare (in the centre of the spot where it first appears).
Away from the centre, the irradiation slowly decreases over
distance up to $R_{\rm{}X}{\approx}h$ from the central point
(with inclination angle of the incident X-rays increasing)
and it drops down fast further out. We start by simplifying 
our analysis and assuming that the irradiation does not show 
any gradient across the spot radius up to $R_{\rm{}X}$. Then it 
drops to zero sharply. Linked with the spatial dependence of 
illumination across the spot surface there should be also 
a corresponding change in the incident angle. However, this incident
angle is about 60$^{\circ}$ at most of the spot surface so 
we assume illumination to be isotropic.  
In this way, we consider the reprocessed radiation as coming from
uniformly radiating circular spots. 
Given the incident radiation flux of a single flare,
$F_{\rm{}inc}$, and its intrinsic spectrum $F_{\rm{}loc}(E)$, 
the reflection is calculated through simple scaling,
\begin{equation}
F_{\rm{}refl}(E) = F_{\rm{}loc}(E) {F_{\rm{}inc} \over F_{\rm{}loc}},
\end{equation}
where $F_{\rm{}loc}$ is the radiation flux assumed in {\sc titan/noar} 
computations. This approach neglects aberration and light bending
in the local irradiation event,
which is roughly correct if the flare height is small in comparison with
the radius in the disk. In this manner, we can connect 
the flare luminosity, $L_{\rm{}i}$, the
incident radiation flux and the size of the spot:
\begin{equation}
L_{\rm{}i} = F_{\rm{}inc}\pi R_{\rm{}X}^2.
\label{eq:one}
\end{equation}

We further assume that all flares occur at the same height above the
disk surface, $h$. This is partially motivated by the fact that the disk 
thickness does not depend on radius in standard, radiation pressure dominated 
Shakura-Sunyaev model, but mainly by simplicity of the assumption.
Therefore, all spots are of the same size ($R_{\rm{}X}$).
Finally, we assume that flares are generated above the disk according 
to uniform random distribution between inner radius, $R_{\rm{}in}$, 
and outer radius, $R_{\rm{}out}$. Gravitational energy
available in the disk depends on radius, and so we cannot expect
all flares to be identical. In the present paper we adopt
$L_{\rm{}i}$ as a function of flare location, $R_{\rm{}i}$, 
in the form
\begin{equation}
L_{\rm{}i}(R_{\rm{}i})\propto
\left({R_{\rm{}i}} \over {R_{\rm{}in}}\right)^{-\beta_{\rm{}rad}}.
\label{eq:Li1} 
\end{equation}
This formula implies that the incident radiation flux 
scales with distance of the flare from the centre as a power law. We
denote the corresponding power-law index $\beta_{\rm{}rad}$
and we normalize the flux amplitude to $F_0$ at the inner rim.
Therefore,
\begin{equation}
F_{\rm{}inc} = F_0 
\left({R_{\rm{}i}} \over {R_{\rm{}in}}\right)^{-\beta_{\rm{}rad}}.
\label{eq:radial1}
\end{equation}

Spots are instantaneously created by the flare occurrence, and hence
one can think of them as being attached to flares and orbiting with 
them at the local Keplerian speed around the central body.
We assume that duration of all flares is the same and equal to
$t_{\rm{}life}$, and we determine the average spectrum as seen by 
the observer after integration over the period $T_{\rm{}obs}$.

In conclusion of this section, we can summarize all nine input parameters
of the model: $M$, $n_{\rm{}mean}$, $R_{\rm{}in}$, $R_{\rm{}out}$, 
$R_{\rm{}X}$, $F_0$, $\beta_{\rm{}rad}$, $t_{\rm{}life}$,
and $T_{\rm{}obs}$. These parameters fully define the flare 
distribution, including the total mean luminosity of the source, 
$L_{\rm{}X}$. In practice it may be more convenient to use 
$L_{\rm{}X}$ as a free parameter instead of, for example, $R_{\rm{}X}$.  

Prediction of the observed spectrum involves two additional
parameters which are inherent to any model involving an accretion disk
around a rotating black hole: inclination angle of the observer, $i$, 
and dimension-less angular momentum of the black hole, $a$. These
parameters span the range $0\degr\leq{i}\leq90\degr$ ($i=0\degr$
the disk axis, while $i=90\degr$ is the disk plane), and $0\leq{a}\leq1$ 
($a=0$ refers to a non-rotating black hole, while $a=1$ corresponds
to a Kerr black hole with maximum rotation). 

\subsection{Properties of the flare distribution}
\label{sect:properties}
In previous section we introduced several assumptions about the flare/spot 
connection. In the present section we use these assumptions to determine 
{\em{}analytical relations describing secondary properties of the flare 
distribution}. Such formulae are very handy for rough estimates and 
for discussion of properties of real sources. Since we assume that the 
spot radius is small in comparison with the overall size of the disk,
also general relativity corrections are assumed to be negligible
for the spot local properties.\footnote{It will be interesting
to relax
this assumption in future work. Also, it should be emphasized that we
do not ignore general relativity effects acting on reprocessed photons
of the spot (see Section~\ref{sec:gr}).}

We first derive the mean luminosity, $L_{\rm{}X}$. Uniform distribution of 
flares over the disk surface means that probability for a flare 
to occur at a given radial and azimuthal position, ($R_{\rm{}i},\phi_{\rm{}i}$),
is equal to
\begin{equation}
p(R_{\rm{}i},\phi_{\rm{}i})\,{\equiv}\,
p(R_{\rm{}i})\,p(\phi_{\rm{}i}) = 
{2 R_{\rm{}i} \over R_{\rm{}out}^2-R_{\rm{}in}^2}\,{\times}\,
{1 \over 2 \pi}.
\label{eq:pr}
\end{equation}
One can associate the average total luminosity of a source with an 
exemplary distribution of flares,
\begin{equation}
L = \sum_{i=1}^{n} L_{\rm{}i}(R_{\rm{}i},\phi_{\rm{}i}).
\end{equation}
The mean (expected) value of this quantity is determined as integral over
the probability distributions,
\begin{eqnarray}
L_{\rm{}X} & =& \sum_{n=1}^{\infty} P(n;n_{\rm{}mean}) \times  \nonumber \\
    &\times&   \int_{\rm{}R_{\rm{}in}}^{R_{\rm{}out}}...\int_{\rm{}R_{\rm{}in}}^{R_{\rm{}out}}
 \int_{0}^{2 \pi}...\int_{0}^{2 \pi} p(R_1)...p(R_n)    \\
   & & p(\phi_1)...p(\phi_n)  \,{\rm{}d}R_1...\,{\rm{}d}R_n \,{\rm{}d}\phi_1...\,{\rm{}d}\phi_n       
    \sum_{i=1}^{n}  L_{\rm{}i}(R_{\rm{}i},\phi_{\rm{}i}),     \nonumber
\label{eq:multi_probab}
\end{eqnarray}
where $P(n;n_{\rm{}mean})$ denotes Poisson's distribution around $n_{\rm{}mean}$.
The multiple integrals can be performed easily, because there is no 
dependence of the integrand on $\phi_{\rm{}i}$. All integrals over $R_{\rm{}i}$ 
are identical (there are $n$ of them) and can be performed independently of each 
other. Summation over the Poisson distribution gives
\begin{equation} 
\sum_{n=1}^{\infty} P(n;n_{\rm{}mean})n = n_{\rm{}mean},
\end{equation}
and the expression for the mean luminosity reduces to
a single integral over the adopted radial distribution of luminosity,
\begin{equation}
L_{\rm{}X} = n_{\rm{}mean} 
\int_{\rm{}R_{\rm{}in}}^{R_{\rm{}out}} p(R_{\rm{}i})L_{\rm{}i} \,{\rm{}d}R_{\rm{}i}.
\label{eq:LX}
\end{equation}
This integral can be calculated analytically. Taking
into account equations (\ref{eq:one})--(\ref{eq:pr}) and 
denoting $\zeta{\equiv}R_{\rm{}in}/R_{\rm{}out}$, we obtain 
\begin{equation}
\int_{R_{\rm{}in}}^{R_{\rm{}out}} p(R_{\rm{}i})L_{\rm{}i} \,{\rm{}d}R_{\rm{}i} = 
 2 \pi F_0 R_{\rm{}X}^2 \; {\zeta^2(1 - \zeta^{\beta_{\rm{}rad} - 2}) \over 
 (\beta_{\rm{}rad} - 2)(1 - \zeta^2)}.
\end{equation}
Using this formula, we can express the spot radius as a function 
of $L_{\rm{}X}$ and other parameters involved:
\begin{equation}
R_{\rm{}X} = \left[{L_{\rm{}X} \over n_{\rm{}mean} 2 \pi F_0} 
{(\beta_{\rm{}rad}-2)(1-\zeta^2) \over 
\zeta^2 (1-\zeta^{\beta_{\rm{}rad}-2})}\right]^{1/2}.
\label{eq:RX}
\end{equation}
At any given moment, the mean covering factor of the disk surface with spots is 
\begin{equation}
c_{\rm{}mean} = n_{\rm{}mean}{R_{\rm{}X}^2 \over R_{\rm{}out}^2 - R_{\rm{}in}^2}.
\end{equation}
The disk surface is completely covered with reprocessing spots
if $c_{\rm{}mean}$ is close to (or greater than) unity.

The mean number of flares during observation of duration 
$T_{\rm{}obs}$ is given by
\begin{equation}
N_{\rm{}mean} = n_{\rm{}mean}\left({T_{\rm{}obs} \over t_{\rm{}life}} + 1 \right).
\label{eq:Ntot_mean}
\end{equation}
Obviously, if the observation is much shorter than the flare duration then
$N_{\rm{}mean}$ is determined just by the mean number of flares. On the
other hand, $N_{\rm{}mean}$ linearly increases with time in the opposite 
limit of a very long observation.

Since some of the flares existing during the observation were born before
the beginning of the observation or they will disappear after the end of
observation, the mean observing time of any single flare is 
clearly shorter than the flare duration time. We therefore introduce effective
life-time, $t_{\rm{}life}^{\rm{}eff}$, which is a combination of the two 
timescales:
\begin{equation}
t_{\rm{}life}^{\rm{}eff} = {T_{\rm{}obs}  \,t_{\rm{}life}\over T_{\rm{}obs} + t_{\rm{}life}}.
\end{equation}

The model specifies only the mean number of flares, but in each specific 
observation we expect various number of flare events to occur. 
They are scattered across the disk and
we model this distribution by drawing the number of flares from
Poisson's distribution around $N_{\rm{}mean}$. To this aim we assume 
all flares last for $t_{\rm{}life}^{\rm{}eff}$.

Therefore, each simulated observation will show clear
dispersion of the total luminosity around the assumed mean value, 
$L_{\rm{}X}$. We can calculate this dispersion analytically,
deriving the mean number of flares during observation $N_{\rm{}mean}$ 
from Eq.~(\ref{eq:Ntot_mean}). Luminosity of a {\em single} flare 
is specified by Eq.~(\ref{eq:one}) and combined with 
Eqs.~(\ref{eq:radial1}) and (\ref{eq:RX}), assuming 
random fluctuations around the mean value of flare number, 
as well as random variations in the distribution of
flares and spots over the disk surface.

The time-integrated luminosity is given by
\begin{equation}
\int_{t=0}^{T_{\rm{}obs}} L(t) \,{\rm{}d}t = 
\sum_{i=1}^{N} L_{\rm{}i}(R_{\rm{}i},\phi_{\rm{}i})\,t_{\rm{}life}^{\rm{}eff},
\end{equation}
where $N$ is a random number drawn from the Poisson distribution 
around $N_{\rm{}mean}$, and $R_{\rm{}i}$ and $\phi_{\rm{}i}$ are 
random numbers, which can be derived
from the probability distribution (\ref{eq:pr}).

The mean (expected) value of this quantity is defined in the same way as
in Eq.~(\ref{eq:multi_probab}). This expression simplifies to
\begin{equation}
\left<\int_{t=0}^{T_{\rm{}obs}} 
 \!L(t) \,{\rm{}d}t \right> = N_{\rm{}mean} \,t_{\rm{}life}^{\rm{}eff}
\int_{R_{\rm{}in}}^{R_{\rm{}out}} 
 \!p(R_{\rm{}i}) L_{\rm{}i}(R_{\rm{}i}) \,{\rm{}d}R_{\rm{}i}, 
\end{equation}
because $t_{\rm{}life}^{\rm{}eff}$ is constant, in accordance with the 
assumptions. It is also evident that
\begin{equation}
\left< \int_{t=0}^{T_{\rm{}obs}} L(t) \,{\rm{}d}t \right > = L_{\rm{}X} T_{\rm{}obs}.
\end{equation}
Variance $\sigma^2$ of this quantity is defined analogously,
\begin{eqnarray}
\sigma^2 & = & \left < \left(\int_{t=0}^{T_{\rm{}obs}} L(t) \,{\rm{}d}t - 
  \left< \int_{t=0}^{T_{\rm{}obs}} 
   L(t) \,{\rm{}d}t \right> \right)^2 \right> \nonumber \\
 & = & \sum_{N=1}^{\infty} P(N;N_{\rm{}mean}) 
\int_{R_{\rm{}in}}^{R_{\rm{}out}}...
  \int_{R_{\rm{}in}}^{R_{\rm{}out}} \int_{0}^{2 \pi}...\int_{0}^{2 \pi} \nonumber \\
  & & \left[ \sum_{i=1}^{N} t_{\rm{}life}^{\rm{}eff} L_{\rm{}i}(R_{\rm{}i},\phi_{\rm{}i}) 
 -  L_{\rm{}X} T_{\rm{}obs} \right]^2  p(R_1)...p(R_N) \nonumber\\
  & & p(\phi_1)...p(\phi_N) \,{\rm{}d}R_1...\,{\rm{}d}R_N \,{\rm{}d}\phi_1...\,{\rm{}d}\phi_N.  
\end{eqnarray}
This integration can be carried out in a similar way as in Eq.~(\ref{eq:multi_probab}),
taking into account the relation
\begin{equation}  
\sum_{N=1}^{\infty} P(N;N_{\rm{}mean}) N^2 = N_{\rm{}mean} (N_{\rm{}mean} + 1). 
\end{equation}
The factor $N^2$ on the left side of this equation comes from the number of mixed terms,
which arise from the expression in square parenthesis. When the square of the first sum 
in square parenthesis is computed, the number of mixed terms is equal to  $N(N-1)/2$.
Therefore, the entire expression reduces to a single integral:
\begin{equation}
\sigma^2 =(t_{\rm{}life}^{\rm{}eff})^2 N_{\rm{}mean}
\int_{R_{\rm{}in}}^{R_{\rm{}out}} p(R_{\rm{}i}) 
L_{\rm{}i}^2 \,{\rm{}d}R_{\rm{}i}.
\label{eq:variance}  
\end{equation}
It is more convenient to use the normalized variance, i.e.\ the
fractional variability amplitude $F_{\rm{}var}$ (see e.g.\ Vaughan et al. 2003) 
\begin{equation}
F_{\rm{}var} = {\sigma^2 \over L_{\rm{}X}^2 T_{\rm{}obs} }
\end{equation} 
which can be expressed as
\begin{equation}
F_{\rm{}var}^2 = {1 \over N_{\rm{}mean}}\;
{\int_{R_{\rm{}in}}^{R_{\rm{}out}} p(R_{\rm{}i}) L_{\rm{}i}^2 \,{\rm{}d}R_{\rm{}i} \over 
\left[\int_{R_{\rm{}in}}^{R_{\rm{}out}} p(R_{\rm{}i}) L_{\rm{}i} \,{\rm{}d}R_{\rm{}i} \right]^2}
\end{equation}
using Eqs.~(\ref{eq:LX}) and (\ref{eq:variance}). In our case this expression is equal
\begin{equation}
F_{\rm{}var} ^2 = {(\beta_{\rm{}rad}-2)^2(1 -\zeta^{2 \beta_{\rm{}rad} -2})(1 - \zeta^2)
\over 2 \zeta^2(2 \beta_{\rm{}rad}-2)(1-\zeta^{\beta_{\rm{}rad}-2})^2}{1 \over N_{\rm{}mean}}.
\label{eq:Nvar}
\end{equation}
The formula is more complex than the simple $1/N_{\rm mean}$ dependence 
due to fact 
that different spots give different contributions to the total luminosity
(parameter $\beta_{\rm{}rad}$ describes the flux energy distribution).
It reduces to the familiar case $F_{\rm{}var} ^2 = 1/N_{\rm{}mean}$ if
all flares have the same luminosity (i.e. $\beta_{\rm{}rad} = 0$). However,
it shows strongly enhanced variability if flare luminosity scales with the 
flare radius proportionally to the dissipative disk flux 
(i.e. $\beta_{\rm{}rad} = 3$) since it this case the formula gives 
approximately  
\begin{equation}
F_{\rm{}var} ^2 \approx {1 \over 8} \left({R_{\rm out} \over R_{\rm in}}\right)^2 {1 \over N_{\rm{}mean}} 
\label{eq:limit}
\end{equation}
for $R_{\rm out}>> R_{\rm in}$.
 
We stress that Eq.~(\ref{eq:Nvar}) for the normalized variance does 
not include relativistic corrections. It can be
applied to the case of a non-rotating Schwarzschild black hole 
with acceptable accuracy, but it turns out to be quite inaccurate for 
maximally rotating Kerr solution.

Naturally, spots should only be generated between the adopted
inner radius of the disk, $R_{\rm{}in}+R_{\rm{}X}$, and the outer radius, 
$R_{\rm{}out}-R_{\rm{}X}$. In order to model a source with a prescribed 
luminosity, non-zero spot size has to be taken into account. A correction
factor is obtained numerically through iterations.
The spot radius $R_{\rm{}X}$ is constrained to be much smaller 
than $R_{\rm{}out}$, and also it must be
smaller than $R_{\rm{}in}$.

\subsection{Generating an exemplary flare/spot distribution}
A specific distribution of spots can be generated from the adopted 
parameters of the model in four steps:
\begin{itemize}
\item[(i)]~choice of the black-hole mass, mean source luminosity, 
duration of observation, and mean properties of the flare 
distribution, i.e.\ input model parameters
$M$, $L_{\rm{}X}$, $T_{\rm{}obs}$, 
$n_{\rm{}mean}$, $R_{\rm{}in}$, $R_{\rm{}out}$, $F_0$, 
$\beta_{\rm{}rad}$, and $t_{\rm{}life}$;
\item[(ii)]~a starting guess of the spot radius $R_{\rm{}X}$ from 
Eq.~(\ref{eq:RX}) and subsequent iterations of the radius, which allows
to reproduce the required luminosity from the flare distribution between
$R_{\rm{}in} + R_{\rm{}X}$ and $R_{\rm{}out} - R_{\rm{}X}$;
\item[(iii)]~generation of the specific number of flares and
corresponding spots, $N$, assuming the Poisson
distribution with the mean $N_{\rm{}mean}$;
\item[(iv)]~generation of the position of each spot from the uniform
random distribution over the disk surface, again within the range 
${\langle}R_{\rm{}in}+R_{\rm{}X},R_{\rm{}out}-R_{\rm{}X}{\rangle}$.
\end{itemize}
The last operation means that the azimuthal angle $\phi_{\rm{}i}$ 
of a flare is determined as
\begin{equation}
\phi_{\rm{}i} = \,{\tt{}rnd}\, \times 2 \pi,
\end{equation}
and the radius $R_{\rm{}i}$ is
\begin{eqnarray}
R_{\rm{}i} &=& [(R_{\rm{}in}+R_{\rm{}X})^2  \nonumber \\
 &+& \,{\tt{}rnd}\, 
 \times\,[(R_{\rm{}out}-R_{\rm{}X})^2 - (R_{\rm{}in}+R_{\rm{}X})^2]]^{1/2},
\end{eqnarray}
where ``${\tt{}rnd}$'' is a random number between 0 and 1.

\begin{figure*}
\epsfxsize=8.8cm 
\epsfbox{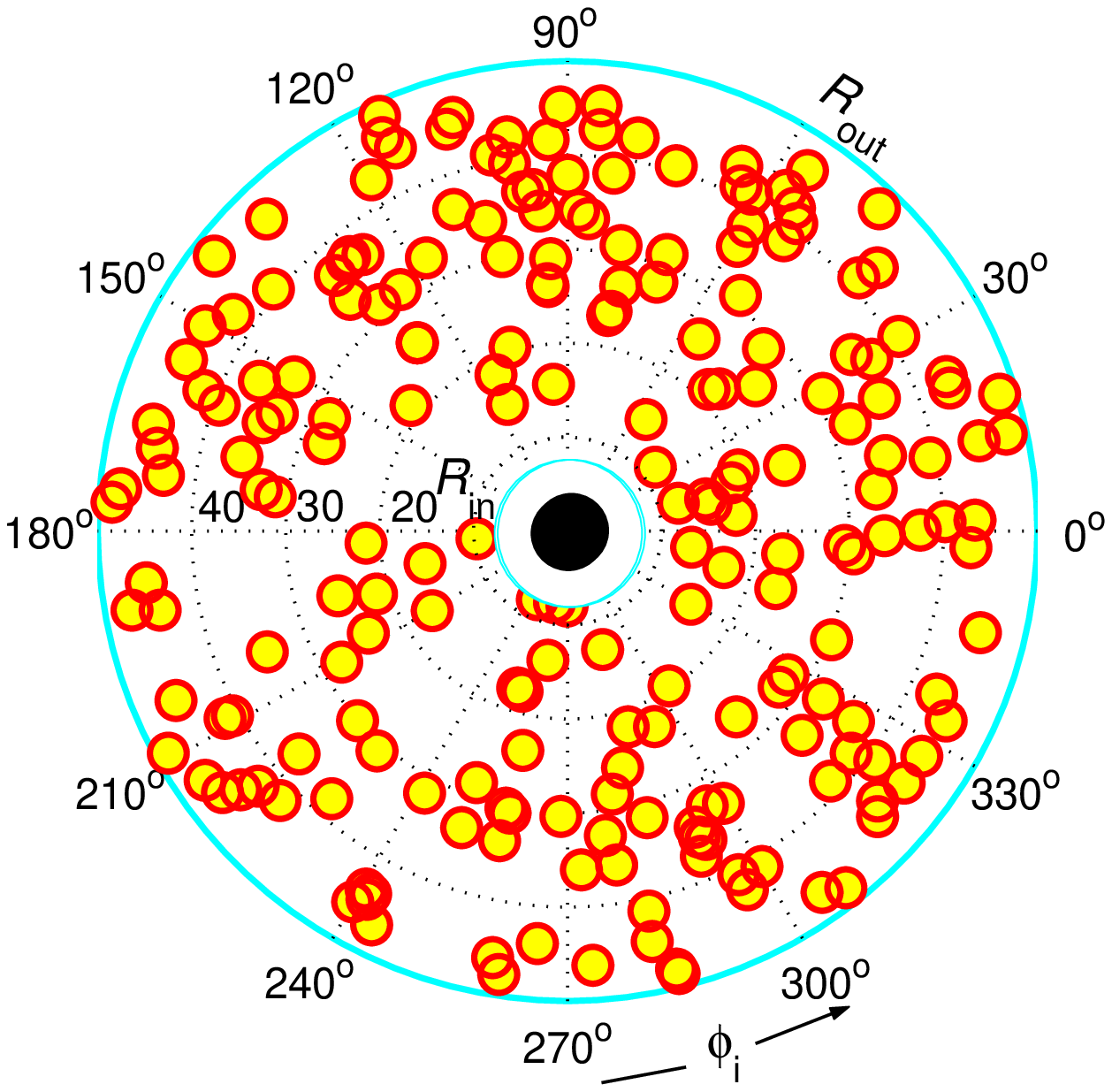}
\hfill
\epsfxsize=8.8cm 
\epsfbox{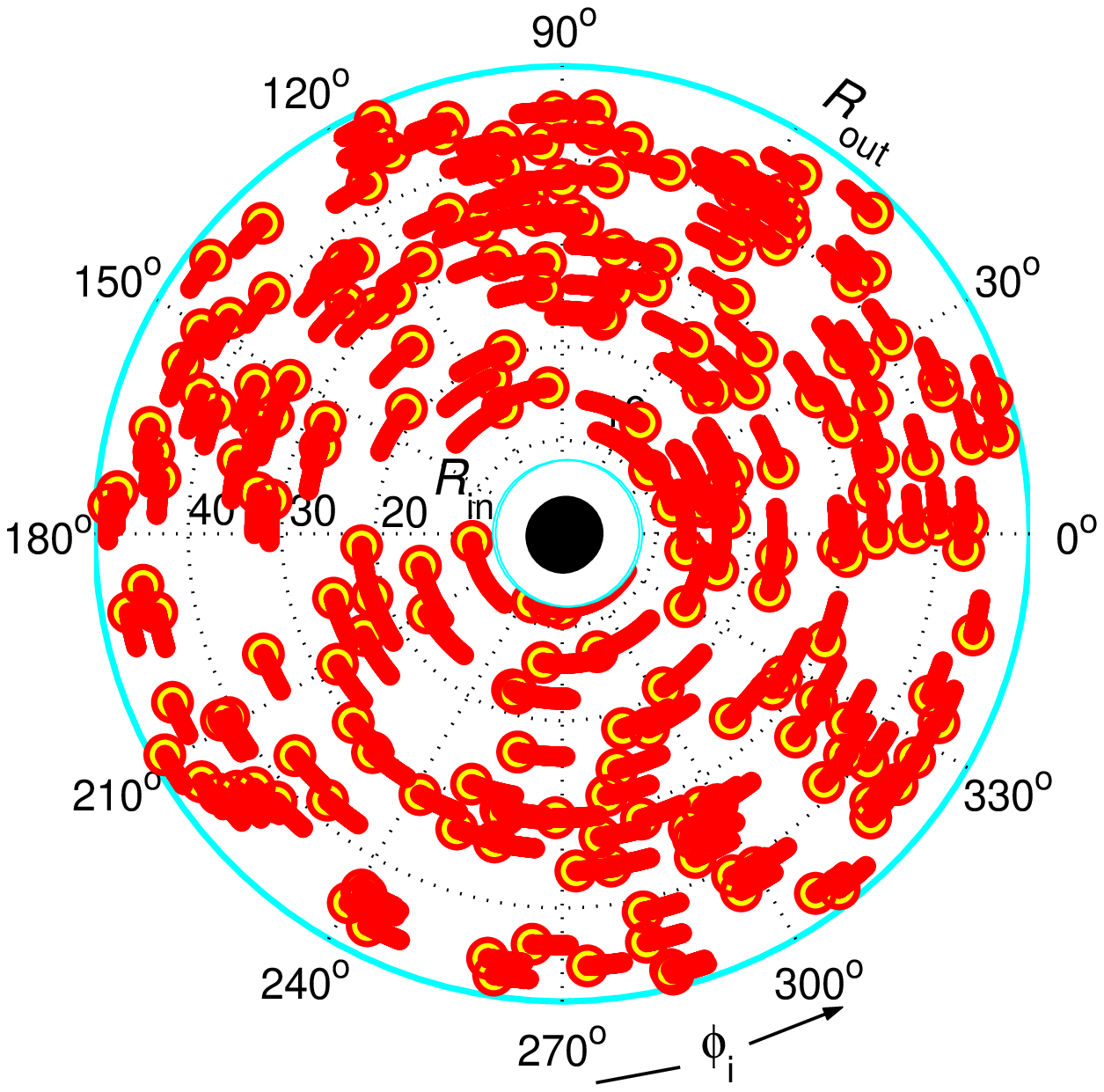}
\caption{Illustration of the spot distribution created by flares which 
illuminate the underlying disk. The plane of the disk is shown, as viewed
from top, along the symmetry axis; $(x,y)$ coordinates are scaled with the black 
hole gravitational radius, $R_{\rm{}g}$. Left: a snapshot of an instantaneous 
spot distribution. Right: effective covering of the disk surface 
by radiating belts in a time integrated observation. Different 
realizations of the spot distribution have been explored in the paper 
(see Table~\ref{tab:properties}).}
\label{fig:spots}
\end{figure*}

An example of the generated spot distribution is shown in Fig.~\ref{fig:spots}.
Each spot is localized at the disk surface at the moment of its 
generation. Spots occurring closer
to the black hole are brighter than those which happen to be born farther out,
because we assumed the power law scaling with $\beta_{\rm{}rad}>0$ 
in Eq.~(\ref{eq:radial1}).

\subsection{Time-integrated properties of spots}
Spots are in Keplerian motion
around the center. During the whole observation each spot moves along
the $\phi$-direction, completing the distance 
\begin{equation}
\Delta \phi(r) = \Omega_K(r) \,t_{\rm{}life}^{\rm{}eff}.
\label{eq:delta_phi}
\end{equation}
Therefore, the integrated emission comes effectively from 
elongated belts rather than circular spots. 
Width of these belts is determined by radius of the 
flare and the length of arc which it has circumscribed, 
$r\,\Delta\phi(r)$.

The belts are relevant for observational properties inferred from the model,
because actual observations have, indeed, some finite (and non-negligible)
duration, $T_{\rm{}obs}$. If the integration time is long enough, some 
spots may complete one whole revolution, or even more
full orbits, and so the length of the belt can be much longer 
than circumference of a circle at the corresponding radius. This is 
particularly relevant for spots in the innermost part of the disk where
motion is fast. Right panel of figure~\ref{fig:spots} shows an example 
of the resulting coverage of the disk surface with radiation belts.

Computation of the time integrated spectrum is therefore performed
as the integration of local intrinsic emissivity of the belts.
The locally emitted flux must be renormalized in order to take into
account that the emission now originates from the belt instead of a
spot. Also, belt size is a function of radius $R_{\rm{}i}$. We find
\begin{equation}
F_{\rm{}belt} =
{\pi R_{\rm{}X}^2 \over  2 R_{\rm{}X}\,R_{\rm{}i}\,\Delta\phi(R_{\rm{}i})}
\;F_{\rm{}spot}.
\end{equation}
The spectral shape of the local spectrum is assumed to be the same
for all spots (see Section~\ref{sect:local_spot}), and the reflected
spectrum is supplemented with the primary emission in the proportion,
which is determined by the radiation transfer computations 
(Sect.~\ref{sect:local_spot}).
The local emissivity is assumed 
to be isotropic for both the primary flare emission and for the
spot/reflected emission. Finally, observed time-integrated spectrum
is subject to the relativistic corrections. 

\subsection{Computations of general relativity effects}
\label{sec:gr}
In order to include general relativity effects in the Kerr spacetime, we
have developed a new computational routine which combines advantages of
different approaches used in the past (Laor 1991; Karas, Vokrouhlick\'y \&
Polnarev 1992; Martocchia, Matt \& Karas 2000). A redesign was desirable
in order achieve sufficient resolution in both energy and time for the
signal arising in multiple spots that are spread over the whole range of
radii. The routine, {\sc ky}, is flexible enough to allow easy modifications 
of local emissivity profiles and it can be used as a rapid stand-alone code as well
as linked to the standard XSPEC package (Arnaud 1996) for X-ray spectral
analysis. In this paper we linked {\sc ky} with the above-described
computations of intrinsic emissivity of the spotted disk.

The routine employs pre-stored data tables to look up and interpolate four
relevant quantities. These are (i)~the energy shift of photons,
(ii)~relative delay of their arrival time, (iii)~magnification of
radiation (lensing effect) in the Kerr metric, and (iv)~the local emission
angle (to account for directional anisotropy of the 
emission).\footnote{In fact, {\sc ky} transfers all four Stokes parameters, 
which are necessary to compute light signal from a polarized source, which
would be received in a detector
equipped with a polarizer. However, polarimetric information is superfluous
for the present paper, and so we ignore it here.} Given the
local (intrinsic) emissivity in the disk plane, the predicted (observed) spectrum can
be integrated. In order to achieve higher accuracy than it was possible
with previous routines, the resolution in energy and time can be controlled
and the grid covering the disk plane can be adjusted. Light rays are
integrated in Kerr ingoing coordinates, so that the frame-dragging effect
is tracked even very close to a fast rotating black hole (Misner, Thorne
\& Wheeler 1973). The equation of geodesic deviation is integrated
separately to ensures precise determination of the flux magnification due
to lensing near caustics.

In general, the disk emissivity is characterized by radiation energy
$F_{\rm em}$ and the related photon flux produced
in the equatorial plane as function of emission energy $E_{\rm em}$,
\begin{equation}
  F_{\rm em}({E}_{\rm em};r_{\rm em},\theta_{\rm em})=
  F(r_{\rm em})\varphi_1({E}_{\rm em})\varphi_2(\mu),
\label{flux1}
\end{equation}
where $F(r_{\rm{}em})$ is total radiation flux emitted at the disk 
surface, $\varphi_1({E}_{\rm{}em})$ is the emissivity profile in frequency,
$\varphi_2(\mu)$ is the limb-darkening law ($\mu$ denotes
the cosine angle between a ray and direction normal to the disk in the
local co-rotating frame). In order to evaluate the terms in
Eq.~(\ref{flux1}), one can employ analytic formulae or tabular data (both
forms are used in
this paper). In the case of
a variable source in Eq.~(\ref{flux1}), time-delay between
light rays is taken into account at the point of their interception
in a detector. Ray-tracing was performed in the Kerr
geometry with redshift function $g$ being given by
\begin{eqnarray}  \label{gfun}
  g & = &
   \frac{\hat{p}_\alpha\hat{\eta}^\alpha}{p_\alpha\eta^\alpha}
   \;=\;g^{tt}\hat{\eta}_t
   +g^{t\phi}\left(\hat{\eta}_\phi-\xi\hat{\eta}_t\right)
     \nonumber \\
   & &
       -g^{\phi\phi}\xi\hat{\eta}_\phi
       +g^{rr}\hat{\eta}_r\hat{p}_r/p_t\,,
\end{eqnarray}
and
\begin{equation}  \label{angle}
  \mu=-\frac{p_{\alpha}n^{\alpha}}
                      {p_{\alpha}\eta^{\alpha}} \; . 
\end{equation} 
Here, $p^\alpha$ and $\eta^\alpha$ denote, respectively, the four-momenta of the
photon and of the emitting material in the disk, and analogously
$\hat{p}^\alpha$ and $\hat{\eta}^\alpha$ that of the photon and of the
observer at $r\rightarrow\infty$; $n^{\alpha}$ denotes a unit space-like
vector perpendicular to the disk surface. For further details, see 
Dov\v{c}iak et al.\ (2003).

As mentioned above, the influence of relativistic corrections on the local spectrum 
can be parameterized by the angular momentum of the black hole, $a$,
and the inclination angle of an observer, $i$. On the other hand, by scaling
lengths with ${R_{\rm{}g}}$ and time intervals with ${R_{\rm{}g}}/c$, one
conveniently ensures that graphs of predicted spectra do not explicitly show 
dependence on the black hole mass.

Relativistic effects, particularly in the case approaching a maximally rotating 
Kerr solution, significantly affect the shape of the radiation spectrum
(especially at large inclination angle). Also the magnitude of observed luminosity 
as well as the observed variance turn out to be more sensitive to inclination
if relativistic effects are taken into account.

\subsection{Fractional variability amplitude}
\begin{figure*}
\epsfxsize=8.8cm 
\epsfbox{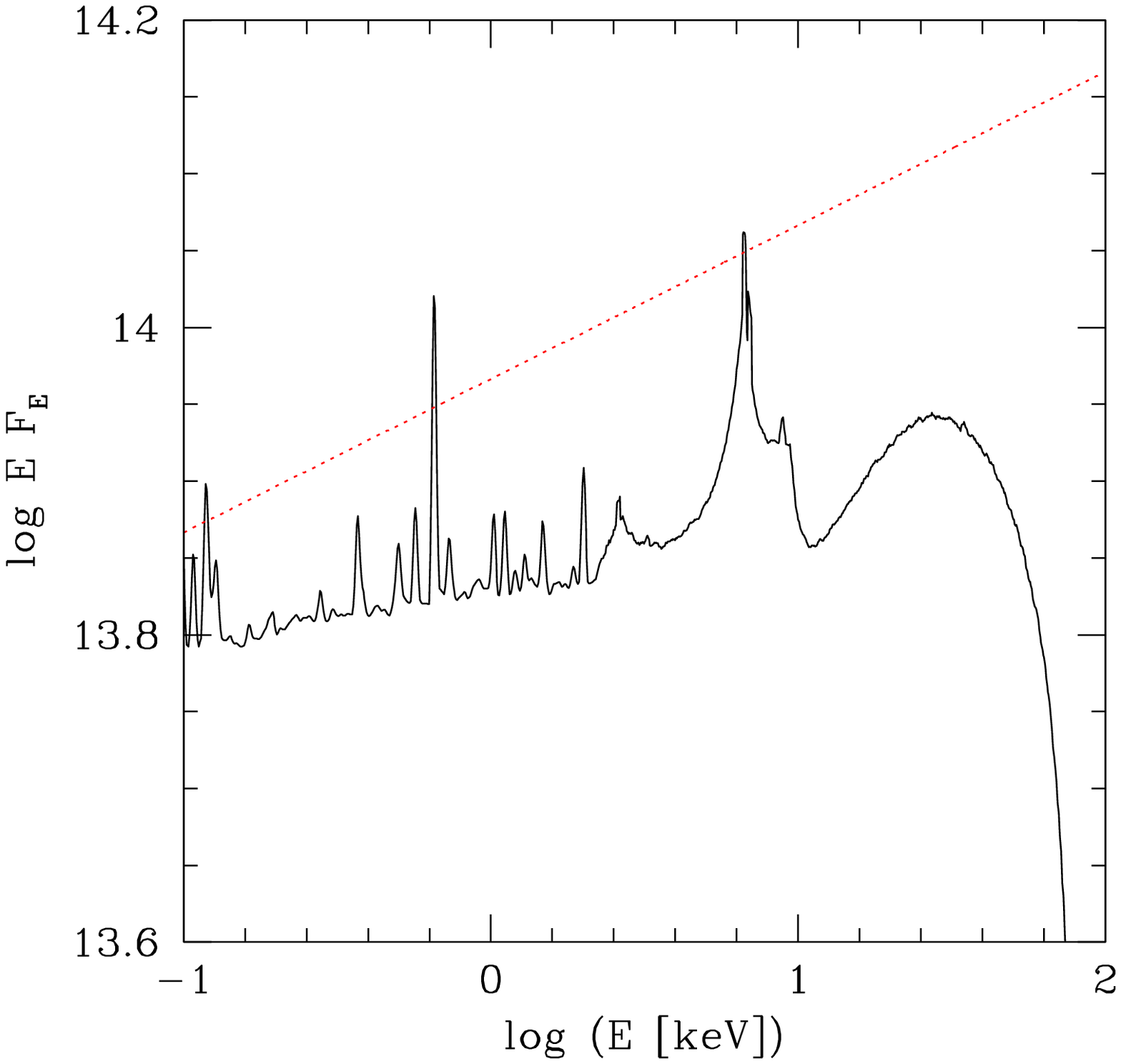}
\hfill
\epsfxsize=8.8cm 
\epsfbox{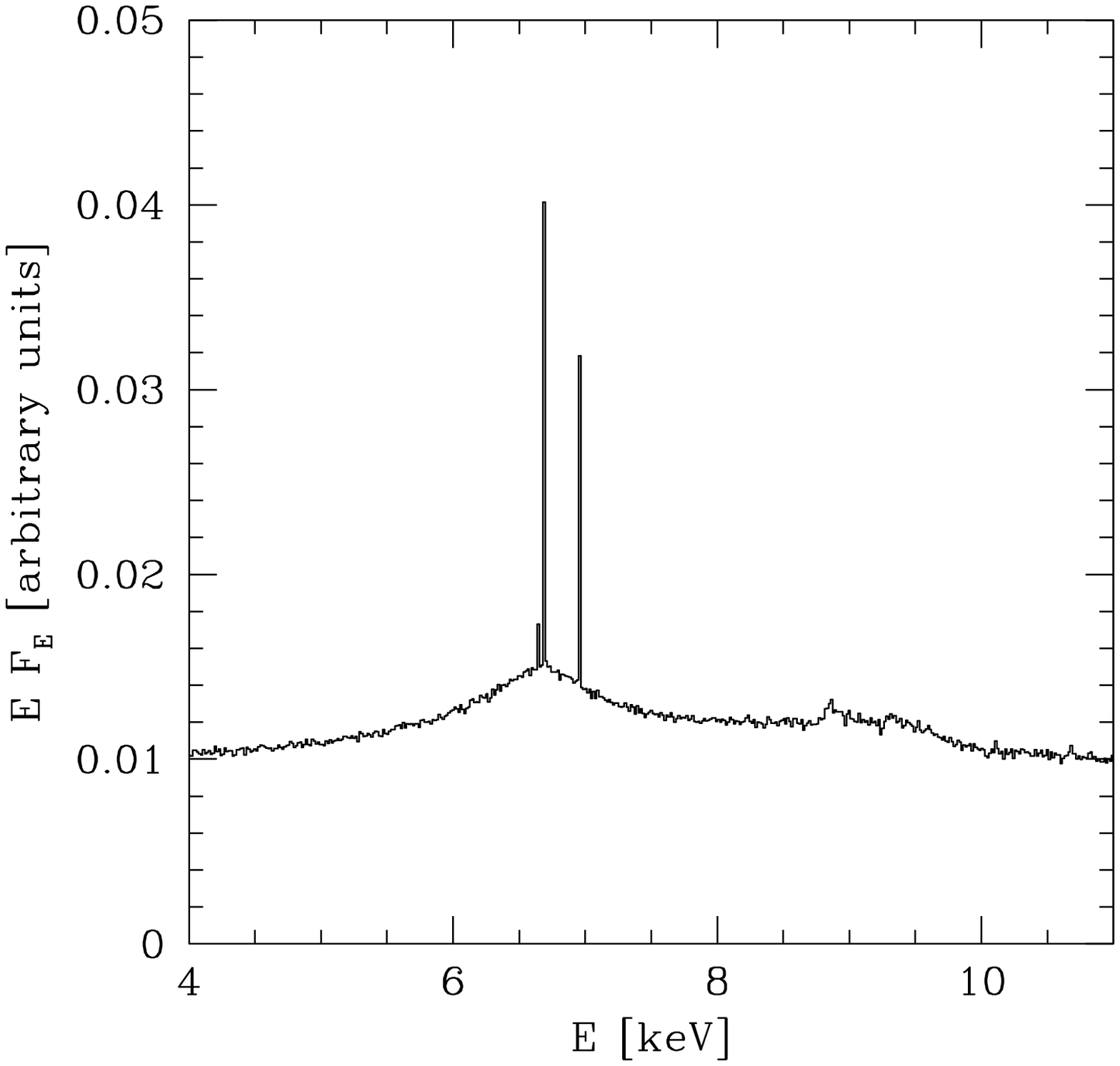}
\caption{Local reflection spectrum of an individual hot spot,
which arises due to irradiation of the
disk surface by a flare. Left: broad energy range is shown 
(log-log scale) with the intrinsic spectral resolution 
$E/\delta{E}=30$. Dotted line shows the incident spectrum.
Right: an expanded part of the same spectrum is plotted around the iron
line energy (linear scale). Here, a higher intrinsic spectral 
resolution was adopted, $E/\delta{E}=100$. In both panels, common 
parameters are $M=10^8 M_{\odot}$, $\dot{m}=0.001$, 
$R=9 R_{\rm{}Schw}$, $F_{\rm{}X}=10^{15}$ erg s$^{-1}$ cm$^{-2}$, $\Gamma=1.9$ 
(see Sect.~\ref{sect:local} for details).}
\label{fig:finspecX}
\end{figure*}

We computed several radiation spectra representing different 
realizations of the same mean spot distribution. In this way we 
simulated independent time-integrated observations of a given source.
We generate each distribution separately of the other, which means that 
we assume the lapse of time among simulated observations to be larger 
than the life-time of flares.

Fractional variability amplitude was then calculated in standard way
(see e.g.\ Edelson et al.\ 2002), as the energy-dependent variance of 
radiation flux normalized by the mean flux at a given energy.
The variance was determined numerically, including relativistic
corrections, from a finite number of realizations, $K$. Then the 
relative error of the normalized dispersion $\sigma$ was estimated as
\begin{equation}
 {\delta \sigma \over \sigma } = \sqrt{\sigma^2 + 1 \over K}. 
\end{equation} 
This dispersion error is purely due to intrinsic statistics properties 
and does not contain any additional finite count rate term usually present in the 
observational data. This specific formula is appropriate for our simulation 
because we draw the number of flares
from Poisson's distribution. Models aimed at reproducing
the entire power spectrum of AGN require more sophisticated approach.

\section{Results}
\label{sect:results}

\subsection{Locally emitted spectrum}
\label{sect:local}

Radiative transfer computations were performed for flare parameters
similar to those used by Ballantyne, Ross \& Fabian (2001). We adopt the black 
hole mass of $10^8 M_{\odot}$. The irradiating X-ray flux, 
$F_{\rm{}X} = 10^{15}$ erg cm$^{-2}$~s$^{-1}$, extends from 1~eV to 100~keV 
in the form of a power law (photon index $\Gamma=1.9$). Iterations between the 
vertical disk structure (in hydrostatic equilibrium) and radiative transfer 
are performed till convergence is achieved (R\'o\.za\'nska et al. 2002). 
We choose the dimensionless accretion rate of $\dot{m}=10^{-3}$,
($\dot{m}$ is scaled to the Eddington rate, 
$2.2 \times 10^{18} M/M_{\odot}$ g s$^{-1}$), and 
radius $r=18 R_{\rm g}$ as typical values where local computations are
carried out. 

Hence, the ratio of the X-ray incident flux to the
disk flux due to the internal dissipation for assumed incident flux 
and accretion rate is equal to 144. Disk thickness is
determined self-consistently by proceeding from the disk
surface down to the equatorial plane and adopting the diffusion approximation 
deeply below the disk atmosphere (for Thomson optical depth 
$\ga 4$). The intrinsic spectrum of a spot (i.e.\ the reflection component) is
shown in Figure~\ref{fig:finspecX}. Notice that the disk surface is 
highly ionized, and so the reflection is efficient also at low energies. 

\subsubsection*{Emission line properties from local spectrum}  
Number of soft X-ray emission lines are visible in spectrum,
despite that the adopted irradiation is very strong.
Since the soft X-ray emission lines have started
to be seen in recent high-quality XMM and 
Chandra data for AGN (Kaastra et al.\ 2002, R\'o\.za\'nska et al.\ 2004), 
we list in Table~\ref{tab:ew} the equivalent widths of the strongest lines, as 
measured with respect to the reflected continuum. The equivalent widths
are determined with respect both to the reflected continuum and to 
the total continuum, i.e.\ primary plus
reflected.

Sulphur and silicate lines are the strongest. Several other lines
from highly ionized species, such as oxygen and carbon, are also well
visible. 

\begin{table}
\caption{Equivalent widths of most intensive 
emission lines from the local spot model (this paper), 
measured (i)~with respect to the reflected continuum, 
EW$_{\rm{}refl}$, and (ii)~to the sum of the 
incident plus reflected continuum, EW$_{\rm{}total}$. Only lines 
with EWs higher than $1$~eV are listed. \label{tab:ew}}
\begin{center}     
\begin{tabular}{llrrr}  
\hline
Ion & Transition & Energy  & EW$_{\rm{}refl}$ & EW$_{\rm{}total}$ \\
    &            & [keV]   & [eV]        & [eV]   \\
\hline
Fe{\sc xix} &               & 0.118 & 1.08  & 0.48\\
C{\sc vi} &$ Ly {\alpha}$   & 0.367 & 1.72  & 0.75 \\
N{\sc vii}& $ Ly {\alpha}$  & 0.500 & 1.01  & 0.44 \\
O{\sc vii} &  f             & 0.568 & 2.59  & 1.11 \\
O{\sc viii}& $Ly {\alpha}$  & 0.653 & 13.32 & 5.69 \\
Fe{\sc xvii}&               & 0.729 & 2.22  & 0.94  \\
Ne{\sc x} & $Ly {\alpha}$   & 1.020 & 4.54  & 1.91  \\
Fe{\sc xxiv}&               & 1.110 & 4.97  & 2.07 \\
Fe{\sc xxvi}& $Ba {\alpha}$ & 1.288 & 2.86  & 1.18 \\
Mg{\sc xii} &$Ly {\alpha}$  & 1.472 & 4.50  & 1.86 \\
Fe{\sc xxiv} &              & 1.495 & 1.48  & 0.61  \\
Fe{\sc xxvi} &$Ba {\beta}$  & 1.739 & 1.26  & 0.51 \\
Si{\sc xiii} & f           & 1.853 & 1.23  & 0.50  \\
Si{\sc xiv} & $Ly {\alpha}$ & 1.999 & 12.02 & 4.95  \\
S{\sc xv} & f              & 2.446 & 1.27  & 0.52  \\
S{\sc xv} & $Ly {\alpha}$   & 2.611 & 13.22 & 5.35  \\
Fe& $ {K {\alpha}}$          & 6.400 &  1.88 & 0.76  \\
Fe{\sc xxv} & i            & 6.630 & 48.69 & 19.75  \\
Fe{\sc xxv} & f            & 6.667 & 37.24 & 15.10  \\
Fe{\sc xxv} & f            & 6.682 & 49.33 & 20.00  \\
Fe{\sc xxv} & r            & 6.700 & 50.07 & 20.30  \\
Fe{\sc xxvi}& $Ly {\alpha}$ & 6.957 & 171.9 & 69.71  \\

\hline     
\end{tabular}
\end{center} 
Two forbidden Fe{\sc xxv} lines are transitions $2p$ $^3P_1^o \rightarrow
1s^2$ $^1S_0$ and
$2p$ $^3P_2^o \rightarrow  1s^2$ $^1S_0$, Fe{\sc xxv} resonance line is
the transition $2p$ $^1P_1^o \rightarrow 1s^2$ $^1S_0$. Two soft X-ray
Fe{\sc xxiv} lines are transitions $3d$ $^2D + 3s$ $^2S \rightarrow  
2p$ $^2P^o$
and $4d$ $^2D + 4s$ $^2S \rightarrow  2p$ $^2P^o$, correspondingly.  
\end{table}

An expanded region of the iron line (in linear scale) is shown in 
the right panel of Fig.~\ref{fig:finspecX}. 
The iron line is strong and multi-component. 
This emission is mostly due to helium-like and hydrogen-like ions. 
The two-component structure
is clearly visible, although $\sim 6.7$~keV component slightly dominates.
Narrow-line components are accompanied by a broad shoulder due to 
Comptonization in the disk surface layers. The iron edge is quite deep, 
as usual in the case of highly ionized medium, and complex, as we see
actually a two separate edges due to Fe{\sc xxv} and Fe{\sc xxvi}, 
at 8.83 and 9.28 keV
correspondingly.
These features are also considerably
smeared by Compton scattering. 

\subsection{General relativity effects in predicted spectra}

\subsubsection*{Relativistically smeared reflection from a disk 
uniformly covered with spots}
\label{sect:uniform}
Thanks to its featureless power-law character,
relativistic effects do not change the spectral shape of primary
continuum emission. However, spectral
lines are still subject to the well-known relativistic smearing 
(e.g.\ Fabian et al.\ 1989; Laor 1991).
It is interesting to see if the smearing of a realistic spectrum 
is distinguishable from the case of separate monochromatic
lines which may be superimposed on broad-band continuum.
To examine the effect of the smearing we consider in this section the
disk surface, which is covered uniformly by a spot-like emission. The local 
emissivity is given by the reflection component, as specified in 
Sect.~\ref{sect:local}. The emissivity is assumed to decrease with 
radius ($\beta_{\rm{}rad}=3$).

\begin{figure}
\epsfxsize=15.0 cm \epsfbox[70 200 490 650]{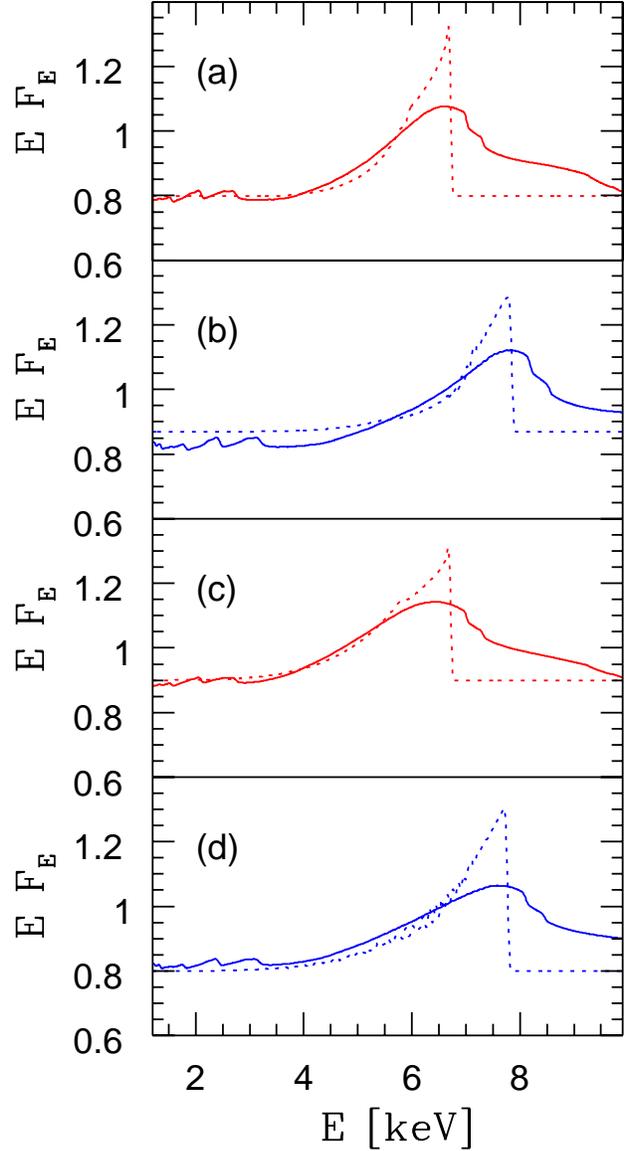}
\caption{The reflection spectrum from the spotted disk.
The local intrinsic emissivity decreases ($\beta_{\rm{}rad}=3$) as radius 
increases up to $R_{\rm{}out}=50 R_{\rm{}g}$. The case of a
Schwarzschild black hole, $a=0$, and moderate inclination, $i=30{\degr}$
(panel a), can be compared with the case $a=0$, $i=60{\degr}$ (panel b),
rapidly rotating case $a=0.998$, $i=30{\degr}$ (panel c), and
$a=0.998$, $i=60{\degr}$.  
Dotted lines show the case of relativistically smeared monochromatic line at
$6.4$~keV superimposed on a flat continuum.}
\label{fig:uni_spec}
\end{figure}

{\it{}There is a clear distinction
between the relativistic smearing of a monochromatic
line and the shape of relativistically smeared realistic reflection 
component.\/} From Figure~\ref{fig:uni_spec} we see that the sharp blue edge, 
which is characteristic for a monochromatic line, is replaced by much smoother 
feature. This is due to the presence of double-peaked
and Compton-broadened iron line in the realistic spectrum, as well
as due to the contribution of the complex and Compton-smeared iron edge.

Weak emission lines of other elements are also noticeable in smeared spectra.
Therefore it is not surprising that traces of these relativistically 
broadened emission lines are
noticeable in the observational data (e.g.\ R\'o\.za\'nska et al.\ 2003 for Ton S180;
Kaastra et al. 2002 for NGC 5548). Soft X-ray lines as strong as those
claimed to be detected in the data by Branduardi-Raymond (2001) and
Mason et al. (2003) are not expected within the frame of our model.

This result clearly supports the argument (\.Zycki et al.\ 1997;
see also Bao et al.\ 1998; Young et al. 1998, Martocchia et al.\ 2000; Gondoin et al.\ 2002; 
Ballantyne et al.\ 2003) that proper interpretation of the observational data
requires consideration of the relativistic broadening of the entire
reflected component, i.e.\ the reflected continuum together with the iron line, 
instead of treating the two components separately.

\subsubsection*{Dependence of the luminosity on the inclination angle}
The importance of the relativistic effects for variability models was discussed
in several papers (e.g.\ Abramowicz \& Bao 1994; Xiong, Wiita \& Bao 2000). 
Its complexity is mostly due to the effect of light bending and Doppler boosting.
Relativistic effects play a crucial role in the observed time-dependent 
behaviour, particularly for the maximally rotating Kerr case.
Therefore, in this section we start by considering all effects 
in this extreme case.

\begin{figure}
\epsfxsize=8.8cm \epsfbox{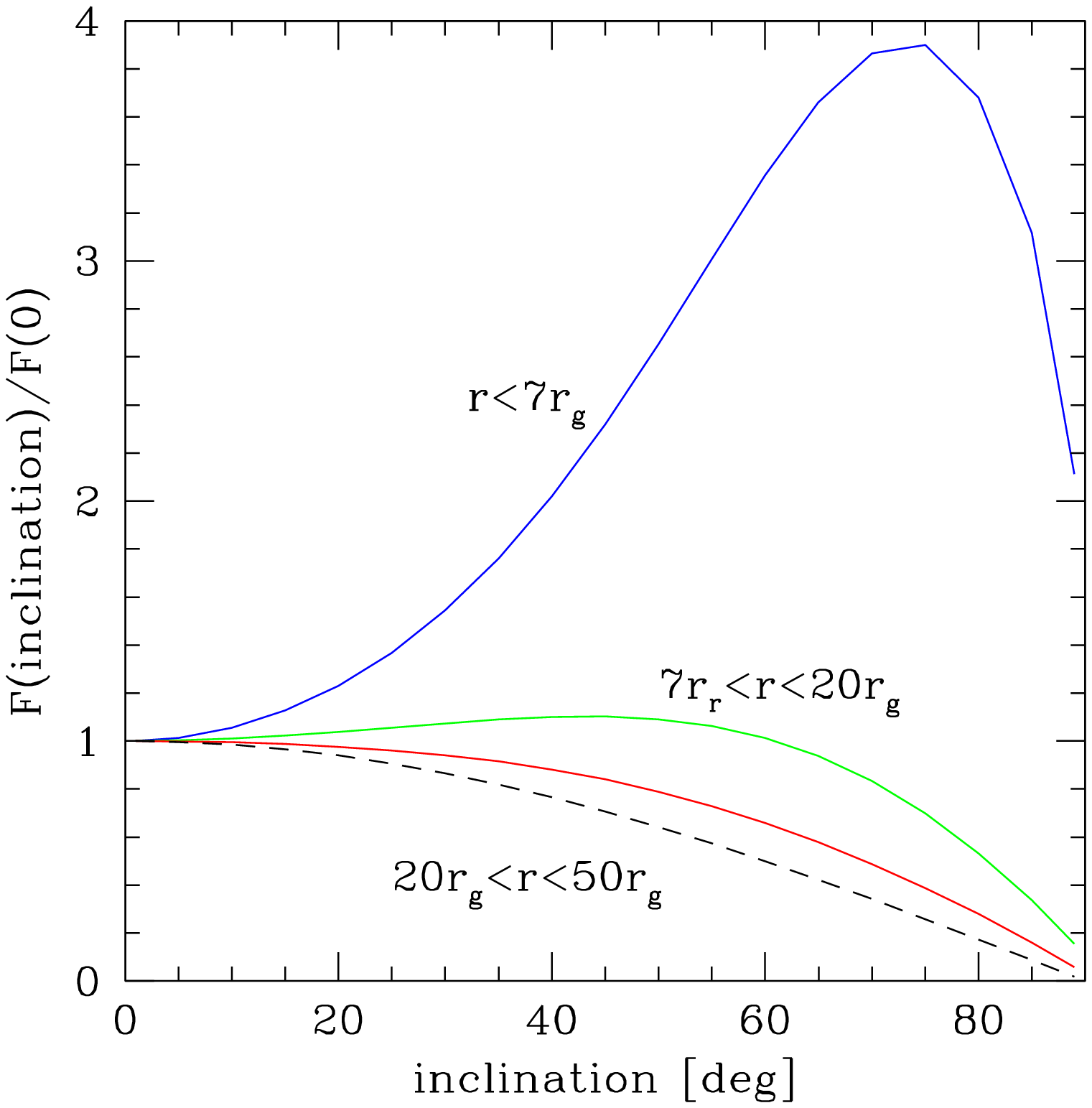}
\caption{The dependence of the luminosity of the uniformly radiating 
disk belts on the inclination angle for the maximally rotating Kerr black hole
$a=0.998$. Power law spectrum and no radial dependence of the 
intensity were assumed.}
\label{fig:incli1}
\end{figure}

The radiation originating in the innermost part of the disk shows
significantly different dependence on the inclination angle than the emission
of the outer parts.
We illustrate this by assuming, as a special example, that the disk surface
is radiating uniformly in the local frame co-rotating with the disk. 
No separate spots and no radial dependence of
luminosity are present. We consider two examples of such spectra: 
(A)~the local emission is a pure power law with an energy index $\alpha = 0.9$,
(B)~the local emission is a pure reflection component as described 
in Sect.~\ref{sect:local}. In both cases, we consider the emission coming from 
three detached regions: (1) an inner ring, $r<7R_{\rm{}g}$, (2) an intermediate ring,
$7R_{\rm{}g}<r<20R_{\rm{}g}$, and (3) an outer ring, $20R_{\rm{}g}<r<50 R_{\rm{}g}$. 
We assume no limb darkening (locally isotropic emission).

In Figure~\ref{fig:incli1} we consider a strictly power-law emissivity
profile and we show the corresponding behaviour of radiation flux at
$6$~keV on the inclination angle. We notice that the emission from the outer 
ring roughly follows purely geometrical effect which (trivial in the flat space), 
i.e.\ the decrease proportionally to $\cos\,i$, where $i$ is the inclination angle 
of observation. However, the emission from the innermost part is more complicated:
moving the observer from the disk axis towards its plane (i.e.\ $i$ increasing), 
the observed flux first brightens, peaks at some intermediate inclination, and
finally it decreases when the disk is inclined very strongly. This behaviour 
is due to competition between the relativistic boosting and reduction
of the projected radiating area. The plot does not depend on energy 
at which the monochromatic luminosity is measured, because the shape of the 
observed spectrum does not depend on inclination in the absence of
spectral features. 

In the case of reflection component the results are practically identical outside
the region of intense iron line. On the other hand, there is more complex 
behaviour within the line range, as can be seen from Figure~\ref{fig:uni_spec}. 

\begin{figure}
\epsfxsize=8.8cm \epsfbox{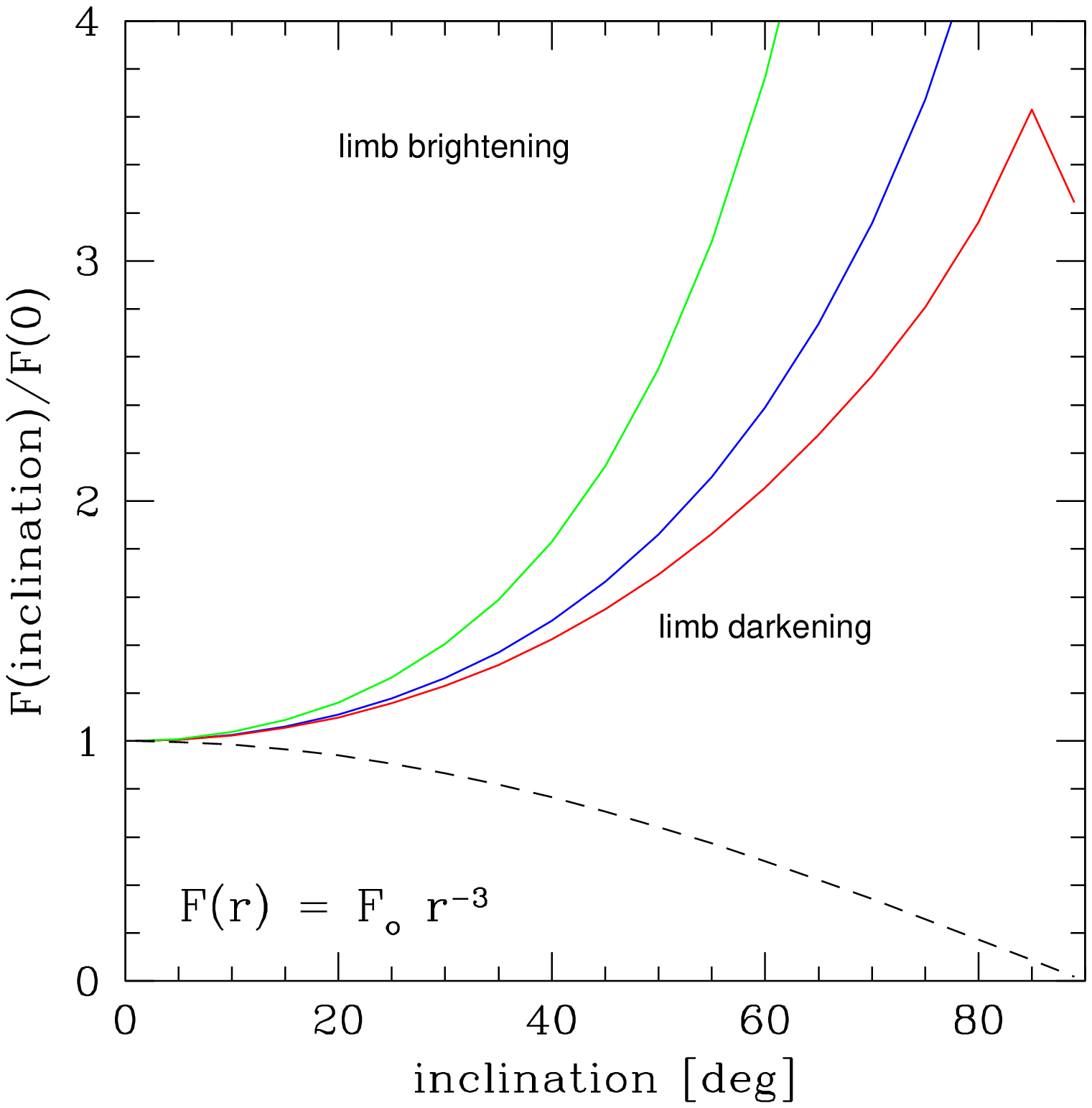}
\caption{The dependence of the luminosity of uniformly radiating disk
on the inclination angle with isotropic local emission, limb brightening 
($I(\mu)\propto5-4\mu$), and limb darkening ($I(\mu)\propto1+2.06\mu$). 
The radial emissivity law $\propto\,r^{-3}$ was assumed together with 
a power-law spectral shape.}
\label{fig:incli2}
\end{figure}

The assumption of radius--independent emissivity (between $R_{\rm{}in}$
and $R_{\rm{}out}$) and zero emission beyond the outer edge is rather 
artificial. Therefore we also show the results for the disk whose radial
emissivity normalization decays with the radius as ${\propto}r^{-3}$
($r\leq50 R_{\rm{}g}$). In Figure~\ref{fig:incli2} observed luminosity 
is plotted as function of inclination. We see that the
dependence is now even stronger than for the emission from 
the inner ring in Fig.~\ref{fig:incli1}, because with the adopted 
steep emissivity law most of the radiation is generated below $3R_{\rm{}g}$. 
Decline in luminosity is observed only at very large inclination, i.e.\ 
close to the disk equatorial plane ($i>85{\degr}$, probably unrealistic 
for real objects). Additionally, at such inclinations very close to 90$^\circ$
multiple images appear (Zakharov \& Repin 2003).

This dependency on inclination angle is only seemingly in contradiction with
several other examples of such trend published in the literature (see e.g.
Gierli\'{n}ski et al.\ 2001, Ebisawa et al.\ 2001, Bhattacharyya et al.\ 2001,
 Ebisawa et al.\ 2001). 
There are three reasons why our results are different. 

First, the emissivity law is usually assumed to be less concentrated towards
the center. When the emission from the multi-black-body 
Keplerian disk around a low mass object is calculated (like in all papers mentioned
above) 
the emissivity goes down to zero at the marginally
stable orbit due to the assumed boundary conditions. This is 
in accordance with the standard thin disk model. However, dissipation in the 
hot corona does not need to follow this law. Also the presence of the large 
scale magnetic field may eventually lead to more dissipation close to 
the marginally stable orbit 
(or even below it) than predicted in models based on zero-torque condition 
(see Afshordi \& Paczy\' nski 2003 for the discussion). 
Models of the broad iron line actually indicate large concentration of the
generated radiation flux towards the center (e.g. Wilms et al. 2001, 
Martocchia et al. 2002).

Second, many computations have assumed some kind of limb darkening.
For illustration we include in Figure~\ref{fig:incli2} the same 
limb-darkening profile as adopted by Laor (1991),
$I \propto 1 + 2.06 \mu $. We see that in this case, under the assumed radial 
emissivity law,  the flux still rises
with the inclination but more slowly. However, limb darkening characteristic
for non-illuminated stellar atmospheres is actually replaced with limb 
brightening in the case of strongly irradiated disk atmosphere.
The angle-dependent computations of the Compton reflection component show
this effect clearly. The increase of emissivity with the inclination strongly
depends on the ionization state of the reflecting material. For the incident
angle of $60{\degr}$ of the incoming beam the ratio of local intensity at 
$90{\degr}$ to that at $0{\degr}$ at $\sim6$~keV is equal to $9$ 
(for weakly ionized 
medium) and $5$ (for strongly ionized medium; cf.\ \.Zycki \& Czerny 1994;
George \& Fabian 1991; Martocchia et al.\ 2000). Therefore, 
in Figure~\ref{fig:incli2} we show also
an example of the flux dependence with limb brightening 
$I(\mu)\propto5-4\mu$ roughly representing
realistic reflection by the strongly ionized medium. In this case
the flux rises almost by $50$\% with the change of inclination angle 
from $0{\degr}$ to $30{\degr}$. However, in the total spectrum the
primary emission dominates over the reflected component. The angular
distribution of this emission is not known so in further considerations
we assume isotropic emission.

The dependence of the observed radiation on the inclination angle has two 
important consequences. First, we recognize the
tendency of the spectral shape on the inclination angle, as already 
illustrated in Figure~\ref{fig:uni_spec}. Second, at low inclination angles
the emission from innermost part of the disk is under-represented in comparison
with the emission from outer parts, while at higher inclinations the opposite
tendency appears. Since in the model outlined in our paper most of the 
variability is due to rare but luminous flares from the inner region, we find
strong dependence of the variability properties on the inclination
angle. We will discuss this in more detail in Sect.~\ref{sect:var_incli}.

\subsection{Fractional variability amplitude}

\subsubsection*{Parameters of the flare distribution}

\begin{table*}
\caption{Exemplary flare distribution properties for the black hole/disk system
seen at a moderate inclination, $i=30{\degr}$. 
Parameters above the double line are assumed while the parameters below the 
double line are calculated as described in Sect.~\ref{sect:flares}.
\label{tab:properties}}
\begin{center}     
\begin{tabular}{llcrrrrrrrrrrrrr}  
\hline
&\multicolumn{1}{c}{\hspace*{-2ex}Parameter} & & \multicolumn{13}{c}{Model\rule[-2ex]{0mm}{5ex}}\\
&                & & A      & B       & C      & D      & E      & F      & G      & I      & J      & K     & L    &  P   &  Q\\
\cline{1-2}\cline{4-16}
\rule{0mm}{4ex}
&$M$             & &$10^8$  &$10^8$   &$10^8$  &$10^8$  &$10^8$  &$10^8$  &$10^8$  &$10^8$  &$10^8$  &$10^8$ &$10^7$&$10^7$&$10^7$\\  
&$R_{\rm{}in}$        & & 6      & 6       &  6     & 1.2    & 6      & 6      & 6      & 6      & 6      & 6     & 1.2  & 1.2  & 1.2\\
&$R_{\rm{}out}$       & & 50     & 50      &  50    & 50     & 100    & 50     & 50     & 50     & 50     & 50    & 50   & 50   & 50\\
&$\beta_{\rm{}rad}$   & & 3      & 3       &  3     & 3      & 3      & 2.5    & 3      & 3      & 4      & 3     & 3    & 3    & 3\\
&$L_{\rm{}X}$         & & 1      & 1       &  1     & 1      & 1      & 1      & 1      & 1      & 1      & 0.1   & 0.5  & 0.5  & 0.5\\
\begin{rotate}{90}Assumed\end{rotate}
&$n_{\rm{}mean}$      & & 30     & 30      & 10     & 100    & 30     & 100    & 30     & 100    & 100    & 30    & 3000 & 1000 & 300\\
&$t_{\rm{}life}$      & & $10^5$ & $10^3$  & $10^5$ & $10^5$ & $10^5$ & $10^5$ & $10^5$ & $10^5$ & $10^5$ & $10^5$&$10^5$&$10^5$&$10^5$\\
&$T_{\rm{}obs}$       & & $10^5$ &  $10^5$ & $10^5$ & $10^5$ & $10^5$ & $10^5$ & $10^4$ & $10^5$ & $10^5$ & $10^5$&$10^3$&$10^3$&$10^3$\\
\hline\hline
\rule{0mm}{4ex}
&$R_{\rm{}X}^{\rm{}a}$   & & 2.65   & 2.65    & 4.59   & 0.62   & 5.16   & 1.57   & 2.65   & 1.45   & 1.16   & 0.84 & 0.24  & 0.43 & 0.80\\
&$R_{\rm{}X}^{\rm{}b}$   & & 3.17   & 3.17    & 6.11   & 0.79   & 7.42   & 1.67   & 3.17   & 1.68   & 1.32   & 0.89 & 0.28  & 0.52 & 1.10\\
&$N_{\rm{}mean}$         & & 60     & 3030    & 20     &200     & 60     & 200    & 33     & 200    & 200    & 60   & 3030  & 1010 & 300\\
&$c_{\rm{}mean}$         & & 0.12   & 0.12    & 0.15   &0.025   & 0.16   & 0.11   & 0.12   & 0.10   & 0.07   & 0.010& 0.09  & 0.11 & 0.14\\
&$F_{\rm{}var}^{\rm{}a}$ & & 43     & 6.0     & 74     &107     & 81     & 18     & 58     & 24     & 34     & 43   & 27    & 47   & 87\\
\begin{rotate}{90}Determined\end{rotate}
&$F_{\rm{}var}^{\rm{}b}$ & & 28     & 4.0     & 38     &64      & 36     & 15     & 38     & 19     & 26     & 37   & 22    & 33   & 45\\
&$F_{\rm{}var}$& & 26       &  3.8       & 39      &27      & 36       & 13       & 36       & 17       & 25       & 32   & 7.0      &  15    &  28 \\
\hline     
\end{tabular}
\end{center} 
$M$ is the black hole mass in $M_{\odot}$; $R_{\rm{}in}$, $R_{\rm{}out}$ and 
$R_{\rm{}X}^{\cdots}$ 
are the disk and the spot characteristic radii in $R_{\rm{}g}=GM/c^2$
(radius $R_{\rm{}X}^{\rm{}a}$ of the spot orbit follows from 
Eq.~(\ref{eq:RX}), while
$R_{\rm{}X}^{\rm{}b}$ includes correction on finite size of spots);
luminosity $L_{\rm{}X}$ is in units of $10^{44}$ 
erg s$^{-1}$ cm$^{-2}$; time intervals $t_{\rm{}life}$, $T_{\rm{}obs}$ are 
in seconds; variances $F_{\rm{}var}^{\cdots}$ are in percent; 
($F_{\rm{}var}^{\rm{}a}$ follows from Eq.~(\ref{eq:Nvar}), 
$F_{\rm{}var}^{\rm{}b}$ includes
the effect of the spot finite size and 
$F_{\rm{}var}^{\rm{}}$ was obtained numerically at 1~keV with 
relativistic corrections and finite size of the spots taken into account).
\end{table*}

The model depends on a number of parameters but most of them are in fact
well constrained by observations, and hence the actual freedom is not so large. 

The mean number of flares at any moment, $n_{\rm{}mean}$, 
cannot be less than $\sim 10$ because observed lightcurves do not show 
strong dips even on short timescales. Neither it can be very large, because
in that case the variability amplitude would be far too small.
In order to quantify variability, we use standard {\sf{}rms} as its measure.
Typical normalized {\sf{}rms} of AGN is known to be of 
order of $0.3$ (see e.g.\ Abrassart \& Czerny
2000; Uttley et al.\ 2002; Markowitz et al.\ 2003). 
The parameter $\beta_{\rm{}rad}$ is expected to be $\sim3$
on the basis of the theoretical arguments as well as successful fits of some
iron $\mbox{K}{\alpha}$ profiles in such sources, including MCG--6-15-30 
(e.g. Iwasawa et al. 1996, Sulentic et al. 1998, Nandra et al. 1997, 
Guainazzi et al. 1999), although more complex distributions were also advocated 
(see Merloni \& Fabian 2003 and the references therein). 

In Table~\ref{tab:properties} we give several combinations of parameters 
that we considered together with the properties of the chosen flare distribution. 
In most cases, we fix the mass of a black hole at a representative value $10^8 M_{\odot}$, 
for which the local computations of the disc structure and radiation
reprocessing were performed. The inclination angle of $30{\degr}$ was
taken as a representative value (the view of the nucleus is frequently obscured 
at high inclinations by the presence of a molecular or dusty torus, so
the average inclination in type~1 objects must be significantly lower than 
$60{\degr}$). The opening angle of the dusty torus is statistically constrained
by the ratio of type 2 to type 1 objects. It was estimated to
be most probably between 2 and 4 (see Krolik 1999; Veron-Cetty \& Veron 
2000). Therefore, the torus opening angle is of order of 
$\sim40{\degr}$, a typical type 1 object is seen at $\sim 30{\degr}$,
and a typical type 2 object at $\sim 70{\degr}$. 

\subsubsection*{Dependence of $F_{\rm{}var}$ on energy}

\begin{figure}
\epsfxsize=8.8cm \epsfbox{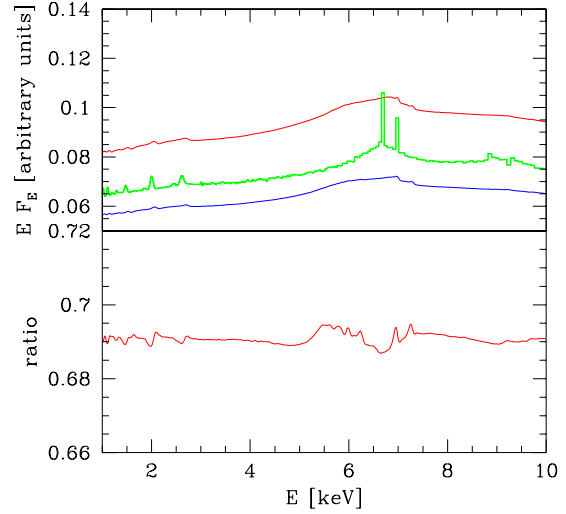}
\caption{Two random realizations of the same spot distribution 
(thin lines) and the original local emission (thick line histogram) 
for model A; the ratio of the two random realizations are shown 
below. The local input spectrum is the sum of the primary (flare)
emission and the reflected (spot) component, without any 
limb-darkening.}
\label{fig:specirat1_2}
\end{figure}

An example of the X-ray spectra from two random realizations of 
model A is shown in Fig.~\ref{fig:specirat1_2}. The intrinsic 
spectrum of spots/flares was assumed to be a combination of the 
intrinsic flare emission and reflected/reprocessed spot emission. 
The observed spectral features are now weaker than the example in 
Fig.~\ref{fig:finspecX} 
where only the spot emission was taken into account. The two 
spectra of statistically identical distributions of spots differ by 
normalization, but otherwise their form is very similar. Variations
are slightly enhanced at energy where the strong iron line occurs. 
Tails of the line vary more significantly because this emission comes 
from several most luminous spots at the innermost part of the disk, 
and so random variations in the position of those few spots
are relatively frequent and important. 

\begin{figure}
\epsfxsize=8.8cm \epsfbox{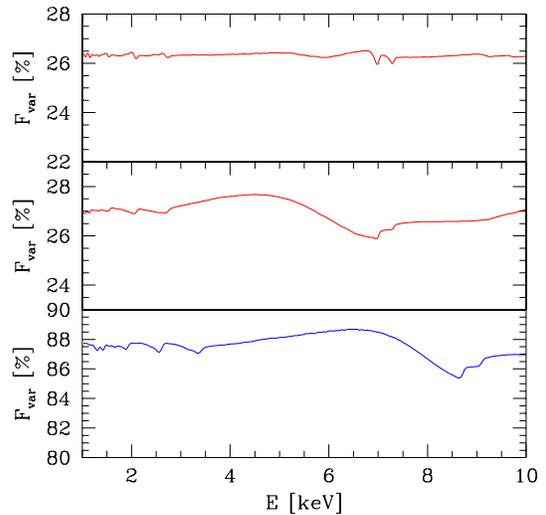}
\caption{Fractional variability amplitude was calculated as a function of energy.
The model A ($i=30^{\circ}$; upper panel), the model D ($i=30^{\circ}$;
middle panel), and the model D at high inclination observed
($i=70^{\circ}$; bottom panel). Duration of a single observation $10^5$~s, 
number of exposures $300$.}
\label{fig:var}
\end{figure}

This effect is best seen in plots where the energy-dependent 
fractional variability amplitude is calculated from many random 
realizations. Such a plot (for model A) is shown in Fig.~\ref{fig:var}. 

In the case of a Schwarzschild black hole,
variations are almost independent of energy. There is only
marginal enhancement of variability in the red wing of the iron 
line, around $5$~keV. There is also a small dip in the plot 
right at $\sim7$~keV. However, the overall energy dependence
is not strong, and so the value of the dispersion at $1$~keV
is quite representative for the entire spectrum (Table~\ref{tab:properties}). 

The energy dependence of $F_{\rm{}var}$ is stronger in the case of 
emission from the disk surrounding a fast-rotating Kerr black hole. 
We show the corresponding plot for model D in the lower panel of 
Fig.~\ref{fig:var}. Here we adopted a higher value of the mean number
of clouds, because otherwise the obtained variance would be 
unrealistically high in the context of AGN. We see in this case strong 
variations in the region of the red wing of the iron line, around 5~keV, 
and again a small dip in the plot right at $\sim7$~keV. Still, the 
overall variability is well characterized by the dispersion at 1~keV, as 
given in the Tab.~\ref{tab:properties}.

The fractional variability amplitude must be in this case calculated with 
the relativistic corrections. 
The effect depends also on the inclination angle. An example for the 
model D but computed
this time at much higher inclination angle of $70{\degr}$ is also shown 
in Fig.~\ref{fig:var}. At such a high inclination, line wings are still
broader than at $30{\degr}$, influencing the entire energy range
between 4 and 8 keV. 

\subsubsection*{Dependence of $F_{\rm{}var}$ on the inclination angle}
\label{sect:var_incli}
\begin{figure}
\epsfxsize=8.8cm \epsfbox{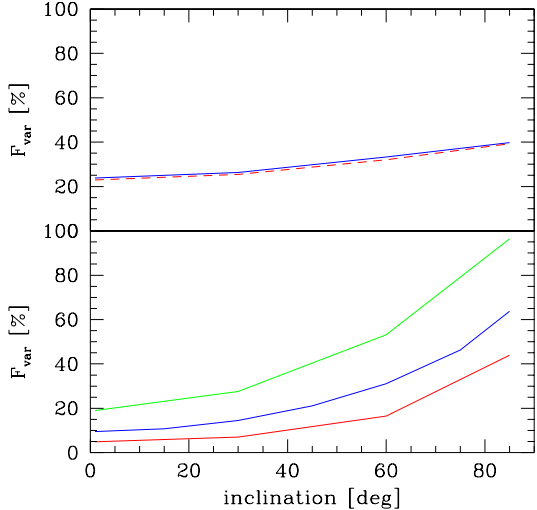}
\caption{The dependence of the normalized dispersion measured at $1$~keV
on the inclination angle of observation for model A (upper panel, continuous line),
model J (upper panel, dashed line) and model L (lower panel, middle curve). 
Two other curves in the lower panel show the result of increasing 
the mean number of flares to $3000$ and decreasing the mean number of flares to $300$. 
Number of exposures was $100$.}
\label{fig:var_incli}
\end{figure}

We have noticed already in Fig.~\ref{fig:var} the strong 
dependence of the overall variance on inclination angle,
which occurs in the case of rapidly rotating Kerr solution. 
In order to examine this trend in more detail,
we plot in Fig.~\ref{fig:var_incli} the variance measured at
$1$~keV as a function of the inclination angle.

There is a noticeable difference between the variability level
of the source observed at $i=0{\degr}$ versus the case of $i=30{\degr}$ 
(the value, which is typically expected for broad-line galaxies on 
the basis of AGN unification scheme). The fractional variability amplitude 
is enhanced by $47$\%. Further increase, by a factor of $3.1$--$4.6$ in the 
variability level, is expected for objects which are observed at
$i\sim60$--$77{\degr}$. Such inclinations are typical for Seyfert~2 galaxies. 
In those objects, in the soft X-ray band, we do not see the direct emission from 
the nucleus. Instead, in some of them the direct emission is visible in hard
X-rays (Done et al.\ 2003), which opens a way to test our prediction
regarding the expected level of variability.

Several author recognized importance of general relativity effects for the 
observed dependence of the variability properties on the inclination. 
Abramowicz \& Bao (1994) noticed the flattening of the power spectrum 
slope with an increase of the inclination angle. Their approximate results were
in general confirmed by Xiong et al.\ (2000) for Schwarzschild geometry. 
Karas (1997) discussed Fourier-phase analysis of the spotted disk model.

The quantitative effect of the inclination depends on details of the 
assumed variability model. We show this dependence for model J, which is
characterized by steeper emissivity ($\beta_{\rm{}rad}=4$) than what 
was adopted in other cases. We see that the rise of the fractional variability
amplitude with the inclination is faster than for $\beta_{\rm{}rad}=3$ 
because more radiation comes from the innermost region where the relativistic 
effects are important.

\subsubsection*{The importance of the limb brightening}
As mentioned in Sect.~\ref{sect:uniform}, limb brightening is expected for
the spot emission. We found, however, that it has no strong effect on the
overall dependence of the fractional variability on the inclination
angle (the increase of the variance with the inclination angle is still
present, although it is slightly reduced). Limb-brightening of the primary 
emission (in the form adopted in Fig.~\ref{fig:incli2}) results in further 
flattening of the 
plots. For the model P, the ratio of the variance
at $60{\degr}$ to the variance at $1{\degr}$ is equal to 
$3.1$ while for the same model
with limb brightening of the spot emission is equal to $2.6$, and for the 
limb brightening of both the flare and spot emission is equal to $2.2$.

\subsubsection*{The accuracy of analytical formula for $F_{\rm{}var}$}
The value of the fractional variability amplitude given by 
Eq.~(\ref{eq:Nvar}) is 
independent both from the energy and inclination angle. In 
Table~\ref{tab:properties} we give these analytical values as
$F_{\rm{}var}^{\rm{}a}$. We also calculate semi-analytical values 
$F_{\rm{}var}^{\rm{}b}$ taking into account the finite size of the spot but 
neglecting relativistic corrections;
a new radius is 
calculated iteratively and inserted into Eq.~(\ref{eq:Nvar}) with
$\zeta = (R_{\rm{}in} + R_{\rm{}X})/(R_{\rm{}out} - R_{\rm{}X})$.

We see that the analytical formula for $F_{\rm{}var}^{\rm{}a}$ 
gives a rather good approximation 
of the exact value in the model K, which involves the Schwarzschild solution
and relatively small size of spots (less than $1\,R_{\rm g}$). 
For other Schwarzschild models
with larger size of spot, the simplest approximation is not 
satisfactory any more, but 
$F_{\rm{}var}^{\rm{}b}$ still provides a reasonable 
approximation. Relativistic effects are
important but the formulae neglecting them can be used for 
qualitative estimates. 

For Kerr models, the analytical value given by Eq.~(\ref{eq:Nvar}) 
over-predicts the 
variability level at low inclination angle by a factor up to 3, but 
it represents 
roughly the correct values at large inclination angles. 
In this case, analytical estimates give
large errors, and so accurate numerical energy-dependent and 
inclination-dependent results 
with general relativity effects are needed. 

\subsubsection*{Trends with assumed model parameters}
We see that short duration of the flare (below a few $\times10^4$~s) leads to
too small normalized dispersion in X-ray lightcurves for a typical
long observation lasting about $1$~day. It does not mean that
short-lasting flares are ruled out, but such flares clearly cannot
be responsible for variability in the timescales of days or more.

Increasing the assumed radius of the disk increases also the variance 
because, in our model, larger $R_{\rm{}out}$ translates
into a smaller number of flares generated in the innermost part of the disk
where flares are the most energetic. This is clearly connected with 
our assumption of the uniform covering of the disk and the power-law radial
decay of flare luminosity, as given by $\beta_{\rm{}rad}$. 
Equivalently, we can modify the 
dispersion by changing the slope $\beta_{\rm{}rad}$: lower value (model F)
reduces variations, while higher value increases it (model J).

\subsection{Special case of MCG--6-30-15}
This nearby ($z = 0.007749$; cf.\ NASA/IPAC Extragalactic Database) 
NLS1 galaxy is one of the best studied objects in the X-ray band.
It is well-known for its 
extremely broad, relativistically smeared iron line (Tanaka et al.\ 1995,
Fabian et al.\ 1995, Iwasawa et al.\ 1996, Fabian \& Vaughan 2003 and the
references therein).
The slope of the power law component in this source 
(photon index $\Gamma=1.87\pm0.01$; Wilms et al.\ 2001) is not significantly 
different from the value $\Gamma = 1.9$ adopted in our radiation transfer
computations.

Energy dependent fractional variability amplitude was studied for this source
by Matsumoto et al.\ (2003) on the basis of ASCA data. They showed that
the overall $2$--$10$~keV amplitude of $\sim 18$--$20$\% is enhanced in the
region of red wing of the iron line (modeled as a broad feature around $5.4$~keV) 
up to $40$\% (but with large error), and somewhat less at its blue wing 
(around $6.4$~keV). Detailed analysis was performed by 
Markowitz et al.\ (2003) on the basis of several RXTE observations.
They found that the overall variability decreases with
energy, going from $\sim20$\% at $4$~keV down to $\sim13$--$14$\% above 
$10$~keV. There is also a slight relative enhancement at $\sim5$~keV and
at $\sim8$~keV, and a dip at $\sim6.5$~keV. Similar trend was noticed by 
Fabian et al.\ (2002) in their long XMM observations,
but in this case the variability level was found to be slightly higher.
We model the observed fractional variability assuming parameters appropriate
for this source. 

The mass of the black hole comes out rather small on the basis of X-ray power 
spectrum analysis: $\log M=6.0$ follows from the normalization method 
(Czerny et al.\ 2001), while $\log M=5.7$ has been derived from the 
high-frequency break model (Uttley et al.\ 2002). However, in view
of uncertainties, the mass as large as $\log M=5.9$ is still compatible. 
No results have been obtained from reverberation studies for this source.
The mass was estimated to be $\log M=7.0$ (Czerny et al.\ 2001)
using the disk method. Reynolds (2000) also argues for this higher 
value on the basis of the bulge--black hole mass relation (Magorian et al.\ 1998)
We thus adopt $M=10^7 M_{\odot}$ in our computations.

X-ray luminosity varies significantly, but taking the typical
$2$--$10$~keV flux to be $\sim 4 \times 10^{-11}$erg s$^{-1}$ cm$^{-2}$ 
(Weaver et al.\ 2001) we can estimate the total X-ray flux ($1$~eV--$100$~keV)
to be of order of $5 \times 10^{43}$ erg s$^{-1}$
(assuming the Hubble constant $H_0=70$ km s$^{-1}$ Mpc$^{-1}$).

The bolometric luminosity of this source can be estimated either from
the $2$--$10$~keV luminosity, taking into account the bolometric correction of 
$27$ (Padovani \& Rafanelli 1998), or directly from the broad-band 
data corrected for extinction. The first method gives 
$1.6 \times 10^{44}$ erg s$^{-1}$ while the second one gives 
$9.1 \times 10^{43}$ erg s$^{-1}$ (Niko{\l}ajuk,
in preparation). This means that for the adopted mass of the black hole 
the source radiates at $\sim10$\% of the Eddington luminosity.


We considered three values of the mean number of flares: $300$, $1000$,
and $3000$. The corresponding models are denoted L, P and Q in 
Table~\ref{tab:properties}. We choose the duration of the observations to
be $10^3$~s, the typical duration of continuous observation
in the data, but the result does not depend strongly on this
value. The adopted life time of flares, $10^5$~s, is chosen
to reproduce the knee in the power spectrum of this source, as determined
by Uttley et al.\ (2002). Therefore, we model the day-to-day variability
(and the result does not apply to longer timescales). 

Comparison of the observed $F_{\rm{}var}$ in short (day-to-day)
timescales (see Figure~5 of Markowitz et al.\ 2003) with theoretical 
level of variability indicates that the most appropriate mean number 
of flares is between $300$ and $1000$.

There is evident similarity in the energy dependence of 
$F_{\rm{}var}$ between observed data and the theoretical diagram:
variability in the region of the iron line red wing 
($\sim 5$~keV) is enhanced, and a dip occurs at about $7$~keV. 
However, the dependence on energy is generally much weaker 
in the model than in the data. Also, the model does not show the
observed strong rise of the $F_{\rm{}var}$ towards low energies. 

This presence of strong variation at $\sim3$--$4$~keV in the data and 
the lack of such variations in the model may indicate that
indeed (Inoue \& Matsumoto 2003), this energy 
band is still affected by varying conditions in the warm absorber.

\section{Discussion}
\label{sect:discussion}

The flare/spot model is an attractive explanation of the X-ray spectra and 
variability of AGN. In this scenario X-ray emission is generated both in hot
magnetic loops above an accretion disks and in the bright spots created under
the loops by strong irradiation. In the present paper we tested this model by 
analyzing the predicted fractional variability amplitude, $F_{\rm{}var}$.

We derive simple analytical formulae which allow to estimate the level
of variability from the assumed mean number of flares, flare duration, 
integration time of a single observation, the ratio of inner to outer disk
radius and the radial dependence of the flare luminosity. This 
energy-independent and inclination-independent formula roughly applies to 
the case of a non-rotating black hole and small size of the spots. The formula is based
on assumption of the uniformly covered disk surface but it can be 
readily generalized
 to another kind of distribution. 

Larger spots and/or fast rotating black hole, with inner disk radius close to
the marginally stable orbit require numerical approach, and we show both the
dependence of the resulting  $F_{\rm{}var}$ on energy and inclination angle.
 
This leads to a firm prediction of our model
which can be used to test the basic scenario. The model shows that 
if the disk extends
close enough to the black hole, the general relativity effects lead to 
significant
enhancement of the variability at large inclination angle of observation.
Therefore, Seyfert~2 galaxies, if intrinsically identical to Seyfert~1 
objects
but viewed at a larger angle, should exhibit statistically higher 
$F_{\rm{}var}$
when measured at the same energy. A factor of 3 difference 
between Seyfert~2 and Seyfert~1 galaxies is expected 
for a Kerr black hole and a factor of 1.4 for a Schwarzschild black hole.

The observational evidence of any trends in this direction is scarce 
at present, but it
may support or reject our view more firmly in future. 
However, some values can be given already now.
Mean variability level of Seyfert 1 objects is $\sim 18$\% 
on the short timescales of 1 day in the sample of Markowitz et al. (2003).
Seyfert 2 galaxy NGC 4945 is seen through the torus and it
displays variability at the level of $\sim 40$\%
in the hard X-ray band. As another example -- 
NGC 7582 (Mihara et al.\ 2000), has
revealed the normalized variability amplitude in the hard X-ray band  
to be about $\sim 30$\%. 
Studies of more Seyfert 2 objects are clearly needed.

The overall intra-day variability level of Seyfert galaxies 
is well explained by the flare model if the mean number
of flares is of the order of 30--100 for assumed non-rotating black hole, 
or 300--1000 for a fast rotating black hole. The number of flares 
requested is much larger than the usual expectation of $\sim 10$ flares.
Such a small number of flares is predicted when an assumption is made that
all flares have the same luminosity. In our more realistic model we allow 
for the flare luminosity to depend on the occurrence radius. When we assume that
the flare luminosity scales with radius as $\propto r^{-3}$, most of the 
source X-ray luminosity comes from a few flares generated in the innermost 
part of the disk which enhances the variability. This trend is clearly seen 
from our analytical expression \ref{eq:Nvar} for the normalized variance,
which, for $\beta_{rad} = 3$, and $R_{\rm out}>> R_{\rm in}$, reduces to 
Eq.~\ref{eq:limit}. Therefore, the total number of flares,
$N_{\rm{}mean}$ can still be quite high for a source with a moderate variance.

The energy dependence of
$F_{\rm{}var}$ is generally weaker in the model than in data. Observed 
variations show trends with energy in 1--10~keV band. For example, in
Markowitz et al. (2003) Akn 564 varies at the level 18--22\%, IC 4329A 
at the level 11--14\% and MCG--6-30-15 at 14--21\%, depending on the
considered energy. In our models the trends in $F_{\rm{}var}$ with 
energy never exceed 2\%. 

This discrepancy can be possibly solved by
relaxing several simplifications: 
\begin{itemize}

\item the local spectrum was computed in detail only at a single radius.
In reality, the shape of the spot spectrum is expected to show 
significant trends with the disk radius, as for example emphasized
by \. Zycki \& R\'o\.za\'nska (2001). A grid of spot spectra should be 
computed (parameters space of the model is rather rich, and so the 
task is numerically very time consuming and we postpone it for future
work); 

\item hydrostatic equilibrium was assumed to compute the irradiated 
disk structure. However, timescales of the flares and
of restoring the hydrostatic equilibrium are roughly comparable (see the 
discussion by Nayakshin \& Kazanas 2002 and Collin et al.\ 2003).
Therefore, neither
the assumption of hydrostatic equilibrium nor the assumption of 
unperturbed disk are satisfactory. Actual disk evolution should be 
followed;

\item we assumed a unique value of flare duration since we aimed at
modeling the variations at the dominant timescale, e.g.\ at the knee of the
power spectrum. A distribution of flare timescales, a coupling between 
the flare occurrence (avalanches), and an exact profile of an
individual flare are needed if 
we want the model to reproduce the entire power spectrum (e.g.\ Lehto 1989,
Abramowicz et al.\ 1991; Xiong et al.\ 2000, Merloni \& Fabian 2001);

\item we neglected the possible effect of the variable warm absorber which 
may be important in the soft X-ray band for some sources, as argued by  
Inoue \& Matsumoto (2003);

\item we neglected the possible contribution of the radiation reprocessed
by some distant reflector, like an outer disk or dusty/molecular torus 
(e.g.\ Krolik et al.\ 1994).

\end{itemize}

It is to be seen whether elimination of these assumptions would lead to 
better agreement of the predicted energy dependence of $F_{\rm{}var}$ 
with the data. Particularly
difficult seems to be the explanation of both apparently lower 
observed variability in the iron line region as well as the enhancement
of the variability towards low energies seen in Fig.~5 of Markowitz et al.
(2003).

We considered just a few special cases along this line. 

A change of the life time
of a flare from radius-independent to scaled with Keplerian timescale, 
without a change of other parameters, resulted in a fractional variability 
amplitude even less dependent on the energy than previously. It is simply 
caused by the fact that in the case of such a scaling relatively more energy
is dissipated in outer region, so the most relativistically broadened and 
variable inner region contributes less to the total lightcurve. The overall 
normalization of the $F_{\rm var}$ depends on the proportionality constant 
between the life times and the Keplerian timescale (with other parameters 
fixed). 

If we assume that the overall radial dependence of the dissipation should not 
be modified, we can consider two representative examples.

In the first case we assume that the life time of a flare scales with the 
Keplerian timescale but the probability of a flare to appear at a given radius
scales inversely with the local Keplerian timescale. We have therefore less
long-living flares in the outer region and more short-living flares in the 
inner region, with scaling of a single flare luminosity with radius unchanged.
In this case  again the fractional variability 
amplitude is less dependent on the energy than in our basic model. Having more 
flares localized in the inner region lead to reduction in the fluctuations in
the relativistically smeared red wing.

In the second case we again assume that the life time of a flare scales with 
the Keplerian timescale, we still adopt the uniform distribution of the flares
across the disk surface but this time we assume that a flare luminosity 
decreases with radius even more strongly ($\beta_{\rm rad} = 4.5$ instead of
usually adopted $\beta_{\rm rad} = 3$) to compensate for an increase of the 
flare life time. Such a solution leads to slightly enhanced dependence of  
the fractional variability amplitude on the energy but the effect is not 
strong.

Complex dependence of the spectral shape of the reflected component on the 
disk radius may introduce significant modification to the predicted energy 
dependence. Clear suggestion of what is needed can be seen from the recent 
analysis of MCG--6-30-15  by Vaughan \& Fabian (2003). A constant
component with complex energy dependence is apparently needed in order to
formally model the fractional variability amplitude in this source (see their
figs. 11 and 16, top panel). This component may perhaps be understood as a
reflection component (above $\sim 1 $ keV) and a contribution from 
emission/scattering by some extended medium (below $ \sim 1 $ keV). However,
it is not clear whether there is any possibility to find a parameter range 
which would satisfy two 'opposing' trends seen in these data: we need strong
reflection from distant disk region in order to reproduce the constant 
component but we need strong reflection from innermost region in order to 
explain the strongly relativistically broadened iron line profile also seen
in these data. Attempts by \Agata \& \. Zycki (2001) and Ballantyne et al. 
(2003) were not successful.

Flares are not the only possibility to explain the X-ray emission of 
AGN. Other scenarios include lamp-post (standing shock) model (e.g. Henri \&
Pelletier 1991, Malzac et al. 1998), model of
gradual or rapid disk evaporation and its replacement by the hot flow 
(e.g. Narayan \& Yi 1994, Liu et al. 2002), possibly with an
outflow (e.g. Blandford \& Begelman 1999), and the
cloud model (e.g. Collin et al. 1996, 
Karas et al. 2000). Further work, taking into account variability issues, is
needed to determine whether the flare/spot model is the most satisfactory.  

\begin{acknowledgements}

We thank Piotr \. Zycki for very helpful discussions and we are grateful to 
Sergei Nayakshin, the referee, for comments which helped to 
improve the manuscript.
Part of this work was supported by grant 2P03D00322 of the Polish
State Committee for Scientific Research and by Jumelage/CNRS No. 16 
``Astronomie France/Pologne''. VK and MD acknowledge support from 
grants GAUK 188/2001 and GACR 205/03/0902 and MD acknowledges 
support from the grant GACR 202/02/0735 in Czech Republic.
\end{acknowledgements}

\end{document}